\documentclass[twocolumn,superscriptaddress]{revtex4-2}\usepackage[]{graphicx}\usepackage[]{color}
\makeatletter
\def\maxwidth{ %
  \ifdim\Gin@nat@width>\linewidth
    \linewidth
  \else
    \Gin@nat@width
  \fi
}
\makeatother

\definecolor{fgcolor}{rgb}{0.345, 0.345, 0.345}

\usepackage{framed}
\makeatletter
 {\par\unskip\endMakeFramed%
 \at@end@of@kframe}
\makeatother

\definecolor{shadecolor}{rgb}{.97, .97, .97}
\definecolor{messagecolor}{rgb}{0, 0, 0}
\definecolor{warningcolor}{rgb}{1, 0, 1}
\definecolor{errorcolor}{rgb}{1, 0, 0}

\usepackage{alltt}\usepackage[]{graphicx}\usepackage[]{color}

\usepackage{soul}
\usepackage{natbib}
\usepackage[skip=2pt,font=small]{caption}
\usepackage{xr-hyper}
\usepackage{hyperref}

\IfFileExists{upquote.sty}{\usepackage{upquote}}{}

\begin{abstract}
    Faculty at prestigious institutions dominate scientific discourse, with the small proportion of researchers at elite universities producing a disproportionate share of all research publications.
    Environmental prestige is known to drive such epistemic disparity, but the mechanisms by which it causes increased faculty productivity remain unknown. 
    Here we combine employment, publication, and federal survey data for 78,802 tenure-track faculty at 262 PhD-granting institutions in the American university system between 2008--2017 
    to show through multiple lines of evidence that the greater availability of funded graduate and postdoctoral labor at more prestigious institutions drives the environmental effect of prestige on productivity.
    In particular, we show that greater environmental prestige 
    leads to larger faculty-led research groups, which drive 
    higher faculty productivity, primarily 
    in disciplines with 
    research group collaboration norms.
    In contrast, we show that productivity does not increase substantially with prestige for either faculty papers published without group members, nor group members themselves.
The disproportionate scientific productivity of elite researchers is thus largely explained by their substantial labor advantage, 
indicating a more limited role for prestige itself in predicting scientific contributions.
\end{abstract}
\IfFileExists{upquote.sty}{\usepackage{upquote}}{}
\begin{document}

\title{Labor advantages drive the greater productivity of faculty at elite universities}

\author{Sam Zhang}
\affiliation{Department of Applied Mathematics, University of Colorado, Boulder CO 80309}
\author{K. Hunter Wapman} 
\affiliation{Department of Computer Science, University of Colorado, Boulder CO 80309}
\author{Daniel B. Larremore}
\affiliation{Department of Computer Science, University of Colorado, Boulder CO 80309}
\affiliation{BioFrontiers Institute, University of Colorado, Boulder CO 80309}
\author{Aaron Clauset}
\affiliation{Department of Computer Science, University of Colorado, Boulder CO 80309}
\affiliation{BioFrontiers Institute, University of Colorado, Boulder CO 80309}
\affiliation{Santa Fe Institute, Santa Fe, NM 87501}

\maketitle

\section{Introduction}
Scientific productivity, crudely quantified by counts of scientific publications, is a basic measure of scientific progress, and its accumulation creates our collective record of scientific knowledge.
However, researchers at more elite universities tend to dominate scientific discourse, via greater scientific productivity~\cite{long1978productivity, way2019productivity}, as well as by greater attention in the form of scientific citations~\cite{cole1968visibility, aaltojarvi2008scientific, morgan2018prestige}, more scientific awards~\cite{cole1974social}, and more trainees that go on to become researchers themselves~\cite{cole1974social, clauset2015systematic}.
These epistemic inequalities in who shapes the scientific literature are ubiquitous, appear early in scientific careers, and tend to persist over time~\cite{long1978productivity, way2019productivity, cole1974social}. Understanding the mechanisms that underlie this prestige-productivity pattern would shed new light on the factors that govern scientific progress and inform efforts to accelerate and diversify technological, biomedical, and scientific discovery.

Epistemic inequalities in science raise complicated questions about their causes and effects.
Do these inequalities facilitate or impede scientific progress? Do they reflect sorting by meritocratic characteristics, such as an individual's skill, effort, or potential? 
Are they driven by biases tied to non-meritocratic characteristics like age and gender, or by non-meritocratic structural factors like 
working environment, social connections, or 
privilege~\cite{morgan2018prestige, way2019productivity}?
Among early-career researchers, 
the greater productivity of elite researchers appears to be caused not 
by their academic pedigree, but rather by 
their working environment:\ more elite institutions tend to provide more productive environments to their researchers~\cite{way2019productivity}.
However, the precise mechanisms by which prestigious environments drive greater productivity remain unknown. 

Prior studies have argued that prestigious working environments can induce greater productivity directly or indirectly through a number of factors.
Prestigious work environments could increase an individual's available research time and hence also productivity, by lowering researcher teaching load, e.g., by hiring non-research-track faculty like adjuncts or teaching professors, or by limiting course or degree enrollments~\cite{dundar1998determinants, graves1982economics}.
Similarly, they could lower researcher service load by employing more administrative staff.
Prestigious universities may incentivize individual productivity via greater compensation~\cite{ehrenberg2003studying, stephan2012economics},
may increase research efficiency via better technological support, or may promote more within-institution collaborations via larger departments~\cite{way2019productivity}.
Or, as we study here, prestigious universities may have more productive or available scientific labor, increasing researcher productivity via collaborations with non-faculty junior researchers.

In the sciences, it is common for a faculty researcher to head a group of non-faculty junior collaborators composed of graduate students, postdocs, and in some cases staff scientists and undergraduate students ~\cite{bozeman2004scientists, beaver2001reflections}.
    By publishing together, all collaborators' publication counts increase with each scientific contribution, and past work has shown that increased collaboration is strongly associated with overall productivity~\cite{lee2005impact}.
    In these disciplines, 
    collaboration leading to coauthorship is a basic 
    aspect of successful mentorship of graduate students and postdoctoral researchers~\cite{bozeman2004scientists,nas2009scientist,alberts2010promoting}.
    However, in practice, collaboration and group norms vary substantially across disciplines and over time, which complicates efforts to estimate scientific 
    labor's effect on scientific productivity~\cite{maher2013factors, kamler2008rethinking, lariviere2012shoulders, shrum2007structures, hagstrom1965scientific, hargens1975patterns, kyvik1994teaching}.
    Past studies have found some correlations between academic labor and faculty productivity, but have also tended to be cross-sectional, based on small sample sizes, focused on individual disciplines, or on only certain types of labor~\cite{ebadi2016boost, shaw2008publication, baker1992evaluation} (but see Ref.~\cite{lariviere2012shoulders}).
    No studies have examined the role of funded scientific labor on faculty group sizes.
    Hence, the extent to which the availability of 
    scientific labor drives disparities in scientific productivity, and how or why such an effect varies across disciplines, is unknown. 

Through multiple lines of evidence, we show that differences in 
scientific labor drive substantial prestige-productivity inequalities, and the scientific dominance of elite universities can be explained by their 
substantial labor advantage over researchers at less prestigious institutions, primarily 
in disciplines where faculty lead and collaborate with a research group.
Our analysis leverages cross-disciplinary, longitudinal data on the education, employment, and publications of 78,802 tenured or tenure-track (TT) faculty spanning 4,492 departments across 25 disciplines in science, engineering, and the social sciences at 262 PhD-granting U.S.-based universities, 
which we combine with researcher-level productivity data encompassing 1.6 million publications from the Web of Science.
We complement these data with institution-discipline-level counts of graduate and postgraduate (non-faculty) researchers~\cite{nsfgss}, 
institutional covariates~\cite{scorecard}, and discipline-specific measures of prestige~\cite{clauset2015systematic}.

First, we show that faculty's annual productivity, measured crudely as their mean publications per year, increases substantially with environmental prestige, with elite researchers being roughly twice as productive as researchers at the least prestigious institutions. We isolate the component of total productivity that could be driven by differences in labor by partitioning each faculty's total productivity into two sources: (i)~group productivity (publications coauthored with non-faculty research group members), and (ii)~individual productivity (all other publications). In disciplines with group collaboration norms, a larger group will tend to drive greater group productivity, but not in disciplines without such norms. We show that in such disciplines, group productivity is substantial and grows with prestige, even as individual group members are no more or less productive. Finally, we show that research labor is highly concentrated within prestigious environments, indicating that elite researchers tend to have larger research groups.

We then test this ``labor advantage'' hypothesis using a series of predictive models, showing that funded labor consistently plays a significant role in predicting productivity and group sizes in disciplines with research group collaboration norms, but not in disciplines that lack these norms. Finally, using a matching experiment on mid-career changes of institution, we show that faculty who move to an environment with more available funded labor tend to have groups that are significantly larger after the move
than those who go to environments with less labor. Taken together, these results identify the environmental mechanism by which prestige drives greater scientific productivity~\cite{way2019productivity, cole1968visibility}, and show that it is the profound labor advantage of elite working environments that allows their scientists to dominate scientific discourse.



\section{Data and Preliminaries}

Isolating the mechanisms by which prestigious environments shape researcher productivity is complicated by substantial variability in publishing patterns and rates across disciplines, institutions, researchers, and even years within a career~\cite{way2017misleading}. 
To span 
these sources of 
variability, we construct 
a comprehensive longitudinal data set of individual researcher productivity, 
encompassing 
1.6 million publications by 78,802 tenured or tenure-track (TT) faculty in 4492 PhD-granting departments in the U.S., across 25 scientific disciplines (see Supporting Information).
Faculty doctoral training and employment information were drawn from a data set of researchers with TT faculty positions 2008--2017, provided by the Academic Analytics Research Center (AARC) under a Data Use Agreement, which we algorithmically matched to full scholarly records as indexed by the Web of Science.
Through these publication data, we extracted and associated all coauthors along with their affiliations 
with each faculty researcher in our data set.
We then classify each 
scientific discipline according to whether it exhibits a research-group norm, in which faculty lead a research group and coauthor publications with its members, or not (see Supporting Information), allowing us to compare productivity patterns across disciplines.

We complement these researcher-level data with institution-level information on 
 academic departments, providing within-discipline estimates of an institution's available scientific labor and environmental prestige.
For each institution and discipline, 
we record departmental counts of graduate and postgraduate researchers by funding source from the 2008--2017 NSF Survey of Graduate Students and Postdoctorates in Science and Engineering~\cite{nsfgss}, and define
``funded researchers'' as 
graduate students on research assistantships, fellowships, or traineeships, 
and all postdocs.
(Unfunded researchers are self-funded graduate students and graduate students on teaching assistantships.)
Environmental prestige scores 
are derived from 
discipline-level faculty hiring networks, in which 
prestige quantifies the ability of an institution to ``place'' its graduates within a given discipline as faculty at other institutions~\cite{wapman2021quantifying, clauset2015systematic}. 
This measure of prestige quantitatively organizes a discipline's collective endorsements of doctoral training programs~\cite{surowiecki2005wisdom, han2003tribal, burris2004academic}, and correlates with authoritative rankings like the US News \&~World Report and the National Research Council rankings~\cite{clauset2015systematic}.
To facilitate cross-disciplinary comparisons, we then divide each discipline's institutions 
into prestige deciles, 
such that each decile contains roughly 
equal numbers of within-discipline faculty, but potentially variable numbers of institutions.

Finally, using the coauthors and affiliations extracted from each faculty researcher's publications, we partition their publications into \textit{group} productivity and \textit{individual} (non-group) productivity, using our census-level data on faculty to identify same-address non-faculty collaborators as likely group members (see Supporting Information).
This division allows us to 
compare individual and group 
productivities between disciplines with different collaboration norms, 
estimate the individual productivity of group members, and 
use propensity-score matching and Poisson regression to evaluate the predictiveness of environmental characteristics, including available scientific labor, on total productivity, group productivity, and group size (see Methods and Materials).

\section{Results}


We examine four main lines of evidence for the central role of scientific labor in driving greater scientific productivity at more prestigious institutions.
First, we 
quantify how (i)~faculty 
group and individual productivities, along with (ii)~individual group member productivities, vary with environmental prestige, using a comparison between 
disciplines with and without group collaboration norms to isolate those norms' causal effect. 
Second, we 
measure the systematic growth of funded scientific labor with prestige, which is a necessary condition for mean group size to grow with prestige. 
Third, we establish a causal link between available 
labor and 
faculty group sizes 
using (i)~a set of models to predict 
group size from 
departmental covariates and (ii)~a 
matched-pair analysis of faculty who move mid-career to environments with more, or less, available labor. 
Fourth, we quantify a systematic relationship between larger faculty group sizes and greater group productivity.

\begin{figure*}[t!]
    \centering
    \includegraphics[width=0.9\linewidth]{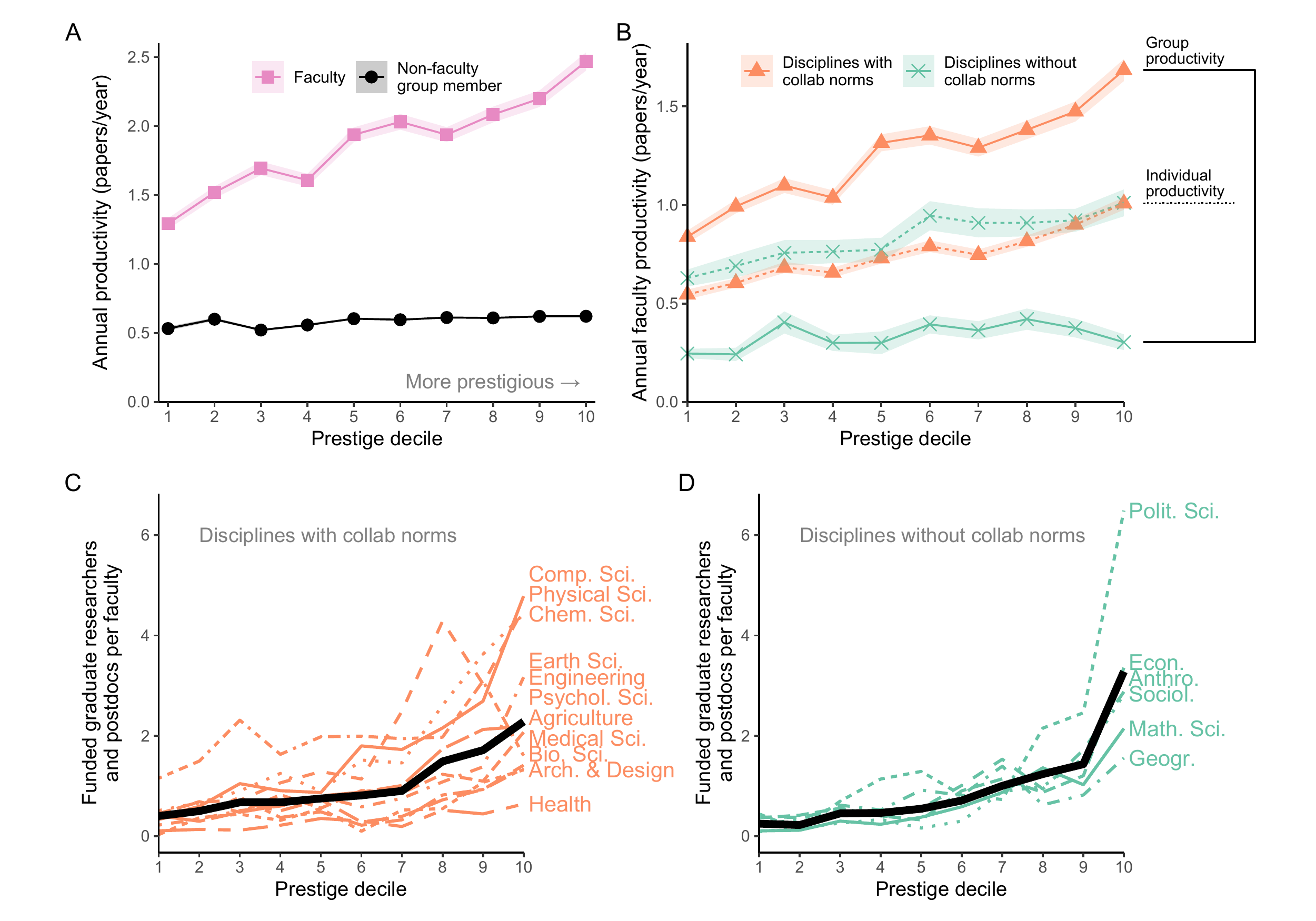}
	\caption{
        Scientific productivity and scientific labor as a function of environmental prestige. 
        (A)~Across all disciplines, average faculty productivity tends to increase with prestige, while average non-faculty 
        productivity does not; higher deciles are more prestigious and shaded intervals denote 95\% confidence intervals.     
        Non-faculty 
        productivity is the total number of papers coauthored 
        by non-faculty group members 
        with faculty, normalized 
        by the length of their collaboration in years 
        (see Supporting Information).
        (B)~A 
        decomposition of 
        faculty total 
        productivity first by 
        group productivity 
        versus 
        individual productivity 
        and then grouped by disciplines with and without research group coauthorship norms (orange and green, respectively),
        showing that individual productivity is similar regardless of collaboration norms, but group productivity is substantially higher in disciplines with collaboration norms.
        (C,D)~Funded scientific labor per faculty, as a function of prestige for disciplines with and without group collaboration norms, showing a systematic labor advantage 
        for the highest prestige institutions regardless of 
        norms; 
        cross-disciplinary mean shown as a thick black line. 
    }
\label{fig:decomp}
\end{figure*}

\subsection{Decomposing the productivity-prestige effect}

First, we 
decompose total productivity, which grows substantially with environmental prestige (Fig.~\ref{fig:decomp}A), into its group and individual components, and then compare these patterns
across disciplines with different coauthorship norms.
If labor drives productivity, then we expect to see both greater total productivity and greater group productivity for faculty working in disciplines where it is normal to lead a research group and share coauthorship with its members.
That is, we expect greater researcher productivity in disciplines with more scientific labor.
Second, we examine the empirical distribution of funded labor across the prestige hierarchy, where 
we expect to see more funded researchers per faculty at more elite institutions.

The 25 disciplines in our data can be divided 
into those with (12) and without (13) group-based coauthorship norms.
Aggregating researchers within each group of disciplines, we find that researchers in the group-norm disciplines publish an average of 1.92 papers per year compared to 1.05 in the non-group-norm disciplines (t-test, $p < 0.001$, Fig.~\ref{fig:decomp}).
However, mean individual productivity in these disciplinary groups is nearly identical (0.74 vs.\ 0.78 papers each year, respectively; t-test, $p < 0.001$).
Hence, the greater productivity of researchers in the group-norm disciplines derives from the excess 
publications they coauthor with group members (Fig.~\ref{fig:decomp}B).

Within both groups of disciplines, individual productivity increases modestly with environmental prestige, 
by only 0.04 additional annual individual publications per prestige decile 
(t-test, $p < 0.001$).
In contrast, group productivity is $1.27$ times greater on average than individual productivity and grows slightly faster, by $0.06$ additional annual group publications per 
prestige decile (t-test, $p < 0.001$).
Hence, in disciplines with research group collaboration norms, 
researcher productivity is predominantly driven by papers coauthored with group members, and group 
productivity correlates strongly with 
environmental prestige (Fig.~\ref{fig:decomp}B).

The strong correlation between 
group productivity and 
prestige could be caused by 
(i)~individual group members becoming more productive, i.e., research group sizes do not covary with environmental prestige but each 
member's individual productivity is higher in more prestigious environments, 
(ii)~research groups becoming 
larger, i.e., the individual group members are no more or less productive 
but group sizes grow with environmental prestige, or (iii)~a mixture thereof.
Across disciplines, we find that 
individual group-member productivity does not increase substantially with prestige, whether we consider disciplines as a group (Fig.~\ref{fig:decomp}A), or separately, e.g., chemical sciences, engineering, biological sciences, and sociology (Fig.~\ref{fig:supp_decomp}). On average, an increase by one prestige decile is associated with a significant but negligible increase of $0.0075$ papers per year for non-faculty (t-test, $p<0.001$).

\subsection{Labor advantages at prestigious institutions}

A necessary condition for the average research group to be larger in a more prestigious environment is that a department's per-faculty available labor must tend to increase with departmental prestige. Holding a funded researcher position typically implies being formally advised by a faculty researcher in the same department---and collaborating with them on research, depending on the discipline's norm. Hence, a department's per-faculty number of funded researchers provides an 
estimate of faculty group sizes in disciplines where faculty lead such groups.

Institution-level counts of funded graduate students and postdoctoral researchers, by discipline \cite{nsfgss}, show that access to funded scientific labor grows with institutional prestige, even as some disciplines employ substantially more labor than others.
With each additional prestige decile, institutions 
gain on 
average 
0.05-1.37 funded graduate and postdoctoral researchers per tenure-track faculty, depending on discipline 
(Figs.~\ref{fig:decomp}C-D,~\ref{fig:supp_absolute_counts}).
This systematic pattern reflects an uneven distribution of labor across prestige (average Gini coefficient $G=0.27$ across disciplines, ranging from 0.11 to 0.36),
and this advantage in available labor at elite institutions appears in all disciplines, not only those with group collaboration norms.

Of the 12 disciplines in our data with group collaboration norms, 9 exhibit a statistically significant increase with prestige in per-faculty funded labor ($p<0.05$, Bonferroni corrected).
The top prestige decile of institutions holds on average 20.2\% of all such funded labor, ranging from 13.2\%~(Computer Science) to a high of 31.6\%~(Biological Sciences).
As a result of this labor concentration, the ratio of funded researchers to faculty in the top decile of institutions is 4.2 times larger, on average, than the ratio in the bottom decile, with the magnitude of this inequality varying substantially by discipline, from a low of 1.4~(Psychology) to 8.2~(Biology) (Fig.~\ref{fig:supp_absolute_counts}).

\subsection{Labor availability drives group productivity}

Together these results suggest a causal relationship in which prestigious environments drive greater faculty productivity by providing more 
scientific labor to individual faculty, via larger research groups.
To test this relationship, we first establish that funded labor availability drives research group size through two complementary analyses:\ 
(i)~we show that funded labor availability explains a significant portion of group size variation in a model of faculty productivity, and
(ii)~we use a matched-pair analysis to show that individual faculty who move to a new environment with more available funded labor tend to form larger groups there than faculty who move to environments with less available funded labor.
Finally, we quantify a systematic relationship between larger 
research group sizes 
and greater group productivity, independent of environmental 
prestige.

\subsubsection{Modeling faculty productivity}
We assess the relationship between 
funded labor availability in a particular environment and individual faculty group sizes by training a series of 
Poisson regression models (see Methods and Materials) to predict 
total productivity, group productivity, and group size in a department from available funded labor, prestige, available unfunded labor, and other departmental 
covariates.
Supporting the labor advantage hypothesis, 
each model shows 
that available funded labor is significant and highly predictive of 
greater total productivity, group productivity, and group sizes in disciplines with collaboration norms (all $p < 0.001$, Fig.~\ref{fig:middle_links}A and Table~\ref{tab:reg}).
On the other hand, in disciplines without collaboration norms, greater funded labor availability does not predict total or group productivity, and is only significantly associated with group size (Fig.~\ref{fig:middle_links}A), a pattern that reinforces the mediating role of collaboration norms on the causal relationship between labor and productivity. 
These findings are robust to alternative model specifications, with 
similar results obtained using
an individual-level Poisson regression with discipline fixed effects and standard errors adjusted for departmental clustering (Fig.~\ref{fig:supp_coefs_ppl}, Table~\ref{tab:ppl_pois_reg}), 
as well as a hierarchical Poisson model on 
individual faculty, with 
departments and disciplines as hierarchical model levels
(Fig.~\ref{fig:supp_coefs_ppl_hierc}, Table~\ref{tab:hierc_pois_reg}).
The latter models indicate a significant role for gender, with men exhibiting both larger productivities and larger groups.
Moreover, we find that funded labor availability predicts increased productivity as the last author, but not as the first author, and only in disciplines with collaboration norms (Table~\ref{tab:reg_firstlast}).


\subsubsection{Matching on re-locations}
If greater funded labor availability 
causes increased faculty productivity,
then 
faculty in disciplines with group-collaboration norms 
who relocate 
from one working environment to another should exhibit a larger productive research group size if the new environment has a larger labor advantage. 
Mid-career moves thus represent a quasi-natural experiment by which to untangle the underlying causal effects of working environment on group sizes, and hence productivity.
We exploit this property using a matched-pair design, in which one mid-career researcher in the pair moves to a working environment with more available labor, while the other moves to an environment with less. 

\begin{figure}[t!]
    \centering
    \includegraphics[width=1\linewidth]{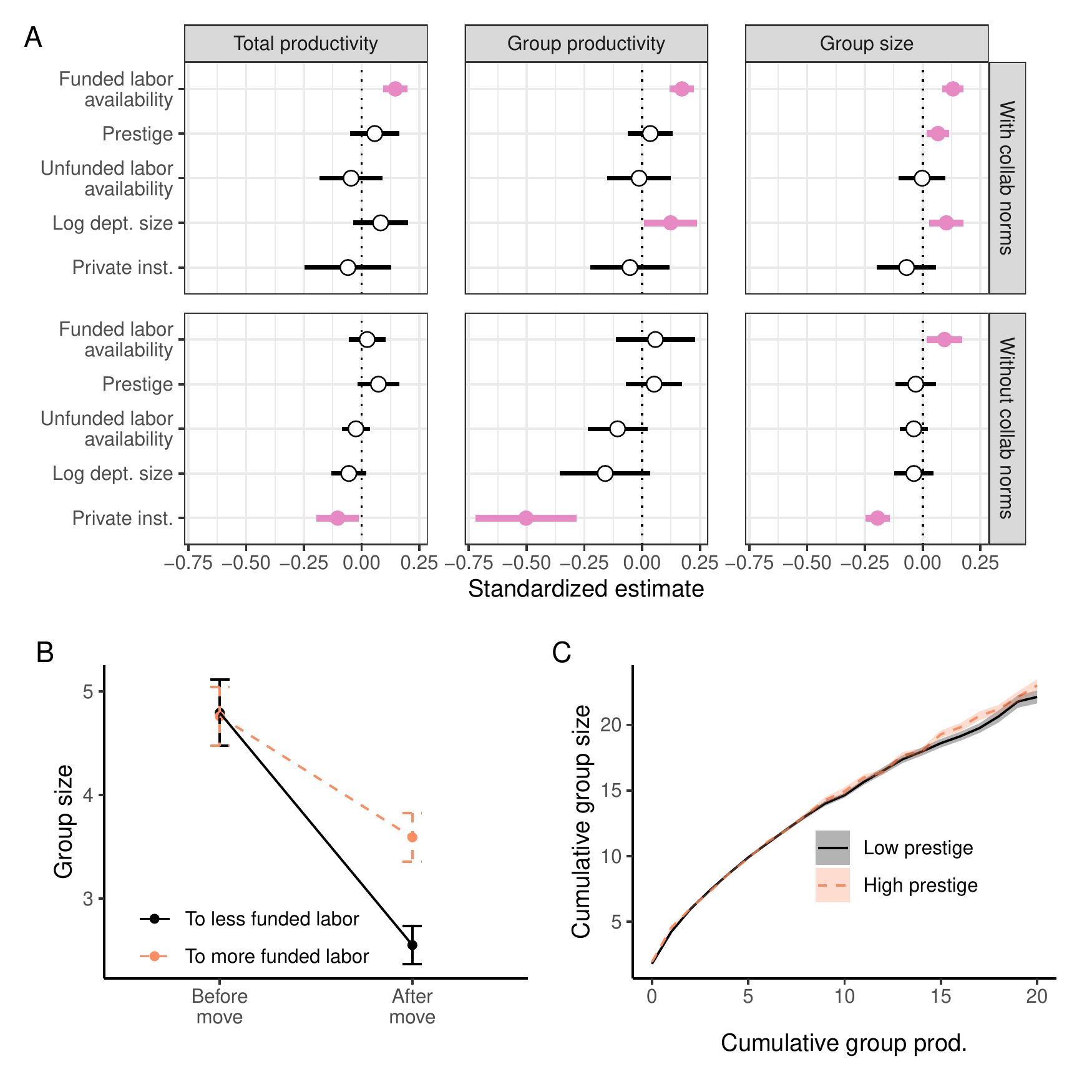}
    \caption{Impact of available labor on group size and group productivity.
        (A)~Coefficients of standardized departmental 
        covariates for 
        predicting 
        total 
        productivity, group productivity, and group size, divided into groups of disciplines with and without collaboration norms (see SI). Bars indicate $95\%$ confidence intervals, and filled-in circles indicate statistically significant coefficients at the $0.05$ level. 
        Funded labor is significant and highly predictive of 
        all dependent variables, even after controlling for prestige, 
        in disciplines with collaboration norms.
        (B)~For matched-pairs of faculty, mean group size in the $3$ years before and after moving to a location 
        with more (dashed orange) or less (solid black) available funded labor than their pre-move location. 
        Error bars indicate one standard error.
        (C)~Mean cumulative number of group members over a faculty career as a function of cumulative group productivity, for faculty 
        at the least prestigious (solid black) or most prestigious (dashed orange) 
        half of 
        institutions in their discipline, 
        showing a nearly identical size-productivity 
        relationship. 
        Envelopes indicate $95\%$ confidence intervals around the means.
    }
	\label{fig:middle_links}
\end{figure}

Matching faculty exactly by discipline and then by full propensity scores derived from other covariates (see Materials and Methods), 
we find 
that faculty who moved to locations with more funded labor developed groups with $0.9\pm0.4$  
more members on average during the third and fourth years after moving than did faculty who moved to locations 
with less funded labor,
which supports the labor advantage hypothesis ($n=778$ faculty; t-test, $p=0.028$; Fig.~\ref{fig:middle_links}B). 
Notably, this effect size is 
averaged across academic disciplines with 
different 
mean group sizes. Hence, we can expect larger effects in fields where research groups are larger, e.g., engineering, chemistry, and computer science.

\subsubsection{Group size and group productivity}

Although individual group members are not substantially more or less productive in high or low prestige environments (Fig.~\ref{fig:decomp}A), it remains possible that
%
group members at more elite institutions tend to work with faculty for longer spans of time, which may confound the apparent prestige-independence of group member productivity. 
    Indeed, we find that group members at more 
    elite institutions tend to have slightly longer 
    productive time spans with faculty (Fig.~\ref{fig:supp_nonfac_min_years}A). However, independent of the number of years used to compute group member 
    annual productivity, the difference across prestige deciles is negligible 
    (Fig.~\ref{fig:supp_nonfac_min_years}B, Fig.~\ref{fig:decomp}A).

Furthermore, if prestige drives group productivity in some way beyond setting the typical research group size, we should expect the cumulative productivity over a faculty career at a prestigious location to increase faster as a function of the cumulative number of group members than for faculty at less prestigious locations. Dividing faculty into those at the most and least 
prestigious halves of all departments in their discipline, 
we find nearly identical cumulative size-productivity curves, indicating essentially no effect for prestige beyond setting group sizes 
(Fig.~\ref{fig:middle_links}C). In fact, these curves are so close that only $5$ of the $20$ values for group productivity are significantly different (t-test, $p < 0.05$, Bonferroni corrected), with a maximum 
significant difference in cumulative group size of only $0.93$, at a cumulative group productivity of
$17$, which has negligible practical significance.

\begin{figure*}[t!]
    \centering
    \includegraphics[width=0.7\linewidth]{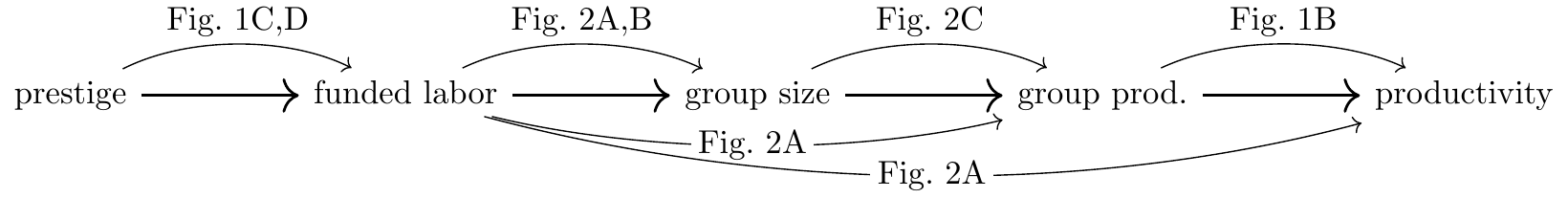}

    \caption{Diagram of causal arguments.
        The 
        association of environmental prestige with greater scientific productivity (Fig.~\ref{fig:decomp}A) is 
        explained by greater 
        available funded labor at elite institutions 
        (Fig.~\ref{fig:decomp}C,D), which 
        drives larger faculty group sizes, even accounting 
        for prestige (Fig.~\ref{fig:middle_links}A,B) and predicts both group productivity and total productivity. 
        Faculty group size itself has a natural and tight relationship with group productivity, 
        independent of prestige (Fig.~\ref{fig:middle_links}C), because 
        group member productivity itself is essentially independent of 
        prestige (Fig.~\ref{fig:decomp}A).
	Finally, increased 
	group productivity can 
	explain the majority of the prestige-productivity effect in disciplines with collaboration norms (Fig.~\ref{fig:decomp}B).
}%
    \label{fig:diagram}
\end{figure*}

\section{Discussion}
A persistent puzzle in understanding 
the drivers of scientific productivity~\cite{shockley1957statistics, hagstrom1965scientific, hargens1975patterns} is identifying why faculty at elite institutions so dominate scientific discourse~\cite{way2019productivity}, producing far more publications than faculty at less prestigious institutions.
Past work~\cite{long1978productivity, way2019productivity} has shown 
evidence that prestigious environments cause greater productivity, at least for early career faculty researchers in Computer Science, but the mechanism for the effect has remained unknown.
%
Using detailed productivity data for more than 78,000 faculty across 25 scientific disciplines, we show through multiple lines of evidence that the greater 
productivity of elite faculty 
can be attributed to 
a substantial labor advantage they hold over faculty at less prestigious institutions (Fig.~\ref{fig:diagram}), which translates into increased faculty productivity in disciplines where faculty lead and coauthor with a group of junior researchers.
Hence, the productivity dominance of researchers at elite institutions is not due to inherent characteristics like greater skill or insight or 
to their academic pedigree~\cite{way2019productivity}, but rather can 
be explained by the greater labor resources accorded to them by their prestigious location 
within the academic system.


By studying 
faculty who change institutions, 
we show that the positive effect on productivity of a researcher's working environment is not limited to the beginning of their career, and instead holds even for mid-career researchers.
The generality of the effect is likely due to the simplicity of its mechanism:\ the creation of scientific knowledge is a collective effort, and hence increasing the number of researchers will reliably increase the amount of research being produced.
Its simplicity suggests that 
sustained increases in available scientific labor for any reason, 
to an individual research group or to an entire institution, can be expected 
to proportionally increase scientific productivity.
The fact that the productivity benefits of 
scientific labor 
accrue primarily to faculty in more prestigious environments illustrates the degree to which funded 
labor is concentrated among those same prestigious institutions (Figs.~\ref{fig:decomp}C-D,~\ref{fig:supp_absolute_counts}).

This unequal distribution of 
scientific labor likely reinforces existing prestige hierarchies~\cite{clauset2015systematic}.
Larger research groups at elite institutions imply more scientific trainees who may themselves aspire to become faculty researchers. 
Hence, the labor advantage may partly explain the dominance of elite institutions in faculty hiring~\cite{clauset2015systematic, wapman2021quantifying}, with all the corresponding implications for influencing research agendas~\cite{morgan2018prestige}, departmental and disciplinary norms, and more, simply because there are many more elite trainees seeking faculty positions than trainees from less elite institutions in any particular faculty search.

The pivotal role of funded labor in explaining the scientific dominance 
of elite institutions also sheds new light on the historical dynamics of the American research ecosystem over the 20th century.
U.S. federal funding for basic and applied research emerged and expanded dramatically over the post-World War 2 
era~\cite{geiger1993research, stephan2012economics}, and historically, elite institutions have received a disproportionate share of those funds~\cite{cole2010great}.
Our results linking funded labor, scientific productivity, and prestige suggest that the post-war funding environment may have enabled a fundamental change in the competitive dynamics of prestige among American universities.
Specifically, external funding enables and encourages faculty-student coauthorship~\cite{maher2013factors}, and the influx of federal funding enabled a coupling between prestige and productivity, allowing institutions to now compete for status by producing a larger number of 
scientific contributions.
The unequal concentration of funded labor among elite institutions then largely reinforced the existing prestige hierarchy despite this new dimension of competition.
However, data on research expenditures and federal funding show a gradual decline in the concentration of resources among elite institutions~\cite{xie2014undemocracy, rouse2018modeling}, suggesting that their ``first-mover advantage'' in the productivity competition may be eroding~\cite{wapman2021quantifying}, as less prestigious institutions increase their share of resources and hence their own scientific productivity.

The lines of evidence described here are derived from observational data, and hence cannot establish causality in the same way a randomized trial might.
As such, the possibility remains that unmeasured variables could account for some of the patterns we observe. 
For instance, there may be a negative relationship between productivity and greater teaching or more administration~\cite{dundar1998determinants, graves1982economics}, and we do find a modest, but unexplained increase in individual productivity associated with more prestigious environments, regardless of disciplinary collaboration norms (Fig.~\ref{fig:decomp}B).
Similarly, labor and productivity interact over the course of an academic career, and are often linked by changes in external funding in the sciences.
Our analysis does not include data on the timing and effect of such funding.
Furthermore, our analyses focus only on publication counts, which are a crude but quantifiable measure of scientific contributions.
As a result, our analysis lacks the capacity to compare individual publications or otherwise assess their intellectual merit or broader impacts. 
Although we find that the productivity of group members does not vary with prestige, other aspects of their scholarship may. 
Elite researchers may tend to work more collaboratively, and we did not directly quantify the effect of such collaborations.
But, we showed that last author publications increased with available funded labor in disciplines with collaboration norms, and there can be at most one last author per paper, suggesting that our results cannot be explained by collaboration effects.
However, past work suggests that many departmental covariates, including doctoral student representation, teaching loads, salaries, and geographic location do not significantly correlate with productivity or in-department collaboration rates~\cite{way2019productivity}.

Elite research environments exhibit a labor advantage in knowledge production because of the substantially unequal distribution of scientific labor across the prestige hierarchy.
A more prestige-equitable distribution of scientific labor is likely to increase the diversity and innovativeness of scientific ideas being explored~\cite{hofstra2020diversity, page2008difference, kozlowski2022intersectional}, in part because what topics are studied varies itself with institutional prestige~\cite{morgan2018prestige, laberge2021subfield}.
That is, increasing the availability of scientific labor at less prestigious institutions may not only change who makes discoveries, but also 
which discoveries are made.
It may also change the relative balance of work on ideas that require large teams and work that is best done by smaller teams~\cite{wu2019large}.

More broadly, our findings have substantial implications for research on the science of science, and in particular for theories of scientific knowledge production that assume meritocratic principles or mechanisms, as these tend to privilege individual characteristics in their explanations and omit environmental or structural mechanisms.
In contrast, our findings on the importance of scientific labor, and its concentration in elite research environments, suggest that individual characteristics and pedigree may play a relatively limited role in certain aspects of knowledge production.
It also suggests relatively simple interventions for both increasing scientific productivity and increasing the diversity of scientific advances.
Accounting for the non-meritocratic effects of research environments will be an important component in developing predictive theories of knowledge production.

\section{Materials and methods}
    The group size for each professor at year $t$ was measured by counting 
unique same-address non-faculty coauthors on their papers for a $3$ year period ending in year $t$. 
This window size captures $94\%$ of productive same-address group members, which 
minimizes the extent to which our measure of productive group size is confounded by productivity (see Supporting Information).
To analyze 
mid-career movements, 
the 806 faculty in disciplines with group-collaboration norms in our dataset who made mid-career moves were divided into those who moved to an environment with more available labor (51.9\%; treatment) and those who moved to an environment with less available labor (control), relative to their starting environment. 
Matching was then performed using 
full propensity scores derived from pre-move covariates of productivity, group productivity, within-discipline prestige, department funded labor-to-faculty ratio, group size, faculty rank, gender, and 
exact matching on discipline. 
The 28 (3.5\%) faculty whose propensity scores were outside the shared region of support between treatment and control groups were dropped, and 
propensity scores for the remaining faculty were recomputed.
Group size in the after-move period was measured as the average size between three and four years after the move, ensuring comparability with the same measurement in the pre-move period.

The full propensity score matching isolates the effect of environmental funded labor on group size and mitigates the confounding effects of differences in academic discipline, faculty rank (assistant, associate, or full), pre-move institutional prestige, pre-move institutional available funded labor, pre-move annual productivity for the four years before the move, and pre-move inferred group size for the four years before the move. The impact of a change in environment on each faculty's productive group size was then measured. Further details are given in the  SI.

A Poisson regression was used to predict 
the averaged departmental productivity using the departmental and institutional covariates given in Table~\ref{tab:reg}. 
The selection of within-discipline institutions 
as the 
unit of regression reflects the fact that both 
the covariates of 
interest and the likely policy interventions 
are environmental rather than individual. 
We report robust standard errors, 
clustered by academic discipline.
Coefficients in the regression are scaled to have zero mean and unit variance.
Funded labor availability is the base-2 logarithm of the ratio of funded researchers (including faculty) to faculty in an institution-discipline.
Unfunded labor availability is the base-2 logarithm of the ratio of unfunded graduate students to faculty in an institution-discipline, and 72 within-discipline institutions 
(9.74\%) that 
had no 
unfunded graduate students in any years were omitted.
All variables were 
averaged over faculty within institution-discipline, then averaged across years.
%


\begin{acknowledgments}
The authors thank Vincent Larivi\`ere and Cassidy R. Sugimoto for helpful comments. Funding:\ This work was supported in part by an Air Force Office of Scientific Research Award FA9550-19-1-0329 (KHW, DBL, AC) and an NSF Graduate Research Fellowship Award DGE 2040434 (SZ).
\end{acknowledgments}

\onecolumngrid
\appendix
\renewcommand{\thefigure}{S\arabic{figure}}
\setcounter{figure}{0}
\renewcommand{\thetable}{S\arabic{table}}
\setcounter{table}{0}

\pagebreak
\section{Data preprocessing}

\subsection{Linking employment records with Web of Science}%
\label{sub:linking_employment_records_with_web_of_science}

For each person in our employment dataset, we first queried their publications in Web of Science (WoS) through the WoS API.
We intentionally created a permissive query that would accept a high false positive rate in exchange for a low false negative rate, since our filtering algorithms could subsequently reduce the false positives but not false negatives.
Furthermore, department names tended to be highly idiosyncratic, and difficult to query exactly.
Thus the query consisted of the last name of the faculty member, the first three letters of the first name followed by a wildcard for the rest of the first name, as well as a standardized name for their institution in the address field that replaced certain conjunctions and punctuation with boolean keywords.

Web of Science returns data on the institution (``org") and departmental unit (``suborg") along with a street address for the address field of authors on papers, and authors are grouped into blocks that share the same address.
Multiple spellings of each org are provided by WoS, with one marked as the preferred entry.
We chose the preferred entry, although it sometimes contained less precise information than some of the alternatives, for example omitting the individual campus for a state university system.
To take advantage of the standardization of the preferred org but to avoid losing geographic data on campus, we created our own organization identifier that concatenated the org with the city.
We found this to be specific enough to match each institution of our employment records directly, and we performed that institutional linkage through a combination of heuristics and manual checking.

We merged the queried publications back to the employment records using the first three letters of the first name, last name, and the mapping that we developed above between the institutions of the two datasets.
We also expanded our result set to include anyone else in the results whose WoS distinct author ID also appeared in the matched set, with the same affiliation as the employment record.
Unlike the preferred org names, suborgs displayed a long tail of unique values, with the same department spelled several different ways. We kept the 10,000 most common suborgs, which included both highly commonly reused names such as ``Dept Biol", but also enough of the long tail.

Next, we computed the most common Web of Science subject associated with papers published by people with affiliations with that suborg name, as long as over 30\% of papers with that suborg label had that subject label.
Each Web of Science author record was associated with a distinct author ID (``daisng ID").
We filtered to daisng IDs who published at least one paper with a matching subject label to their suborg.
This resulted in 1,829,326 papers, each with at least one of 134,776 unique tenure-track faculty authors linked to our employment dataset.
We further have 654,866 WoS distinct author IDs for non-tenure track collaborators on these papers.
Viewing this as a bipartite network where half of the nodes are papers and the other half are authors, with an edge between a paper and an author if that author coauthored the paper, we have 4,932,471 total edges.
We use the employment records, which contain the year that each person received their degree, to filter the data to publications after the year of each person's degree.
This left 1,789,333 publications and 133,747 unique tenure-track faculty before joining with the NSF Survey of Graduate Students and Postdoctorates in Science and Engineering.

\subsection{Linking employment records with labor availability data}
\label{sub:linking_employment_records_with_nsf}

We measure institutional-discipline and departmental counts of funded and unfunded labor through the NSF Survey of Graduate Students and Postdoctorates in Science and Engineering~\cite{nsfgss}, which is an annual census administered to US academic institutions that grant research-based master's and doctoral degrees in science, engineering, and certain health fields.
The census is administered in the fall of the survey year, and has been continuously collected since 1966.
Eligible institutions are determined through the federal Integrated Postsecondary Education Data System (IPEDS), and institutional coordinators collect the data from units within their institutions to report to the NSF.
In 2020, 96.8\% of units provided complete or partial data, and 94.9\% of institutions responded~\cite{nsfgss}.
Missing data are imputed by the NSF.

As funded research labor, we counted all postdoctoral researchers and graduate students on a research assistantship, fellowship, or traineeship.
As unfunded labor, we counted the remaining graduate students, which included teaching assistants and self-funded students.
Prior to 2017, the data did not distinguish between master's and doctoral students.
In 2017, two versions of the data were available, one which separated master's and doctoral students, one which did not.
We consistently used the version of the data that did not distinguish between master's and doctoral students, since our data spanned 2008--2017.
We matched the data at the annual level for all of our employment records, and we averaged the statistics across time at the same time that we did so for other departmental and individual characteristics like counts of tenure-track faculty and productivities.

The data are provided with institution identifiers and codes identifying the field of study.
We manually match each institution identifier with their corresponding identifier in the employment data, where possible, as well as matching the field codes with the disciplinary codes in the employment data.
There could be many units within an institution with the same disciplinary label, and thus we perform a robustness test by analyzing the data at two levels: first, by only including the pairs of units where the NSF unit appears only once, since the NSF unit is at a higher level of aggregation (N=739 units), and second, by accepting all matches and aggregating units up to the institution-discipline level (N=1800 institution-discipline pairs).
For the departmental regression models, we use the strict linkage, but we show that the results are robust to the more permissive linkage in Table~\ref{tab:reg_expansive}.
Since there are few mid-career changes of institution, we use the more permissive linkage to perform the matching experiment to maximize statistical power.
We also use the more permissive linkage for the aggregated statistics that we compute for each prestige decile.

The main quantity of interest is the ratio of either funded or unfunded researchers to faculty in a department.
We would like to take the logarithm of this ratio, but since there are departments with no funded graduate or postdoctoral researchers, this logarithm is improper without some smoothing.
We accomplish this by including the tenure-track faculty themselves as funded researchers, so the funded faculty to labor ratio consists of the ratio of funded graduate and postdoctoral researchers plus the number of tenure-track faculty to the number of tenure-track faculty.

\subsection{Inferring research group members from data}%
\label{sub:inferring_research_groups_from_data}

Research group sizes are difficult to measure accurately, even from seemingly ideal data sources like surveys and CVs.
For example, for faculty with large research groups, students who pass through their lab for a rotation or who switch advisors halfway through their degrees may end up being omitted from that faculty's memory, in the case of a single retrospective survey, and be omitted from that faculty's CV.
Moreover, there is no standard convention for including students and their names in faculty CVs, and when they are included, often only completion dates are included, if dates are included at all.
Approximating the group size is useful as a control variable in our analysis, especially in the matching analysis, where we would like matched sets of people to have similar pre-move group sizes.
Moreover, identifying potential group members at the individual level allows us to quantify their average productivities.

Further, since our concern is with the ways that students contribute to productivity, we seek to further restrict our attention to the active or productive research group of a faculty.
For example, a graduate student who graduates with a masters degree without any publications does not contribute to the productivity of any faculty, and is thus not a part of anyone's active research group.

This criteria, combined with the above difficulties on measuring group size, lead us to infer group members from publication data.
Our heuristic is that a research group member is someone who is not tenure-track faculty, but who coauthors papers with a given faculty member using the same departmental affiliation.
We use the Web of Science distinct author identification system ID to find all of the papers that any given group member coauthors in collaboration with a tenure-track faculty in their department.
This includes funded graduate students from the same department, as well as postdocs, but also any unfunded graduate students on teaching assistantships or undergraduate students, as well as non-tenure-track faculty and research staff who coauthor papers with tenure-track faculty in their departments.

We compute the year span that a group member appears in the data, by subtracting the year of the last publication that they appear in a department from the year of the first publication, plus one.
Since we are interested in the role of graduate and postdoctoral labor, rather than research staff or non-tenure-track faculty, we filter the non-faculty data to exclude people whose year span is 8 years or greater, and anyone whose first publication in a department is before 2007.
We also exclude papers that group members author without any tenure-track faculty coauthors in their department.

\begin{figure*}[ht]
    \centering
    \includegraphics[width=0.7\linewidth]{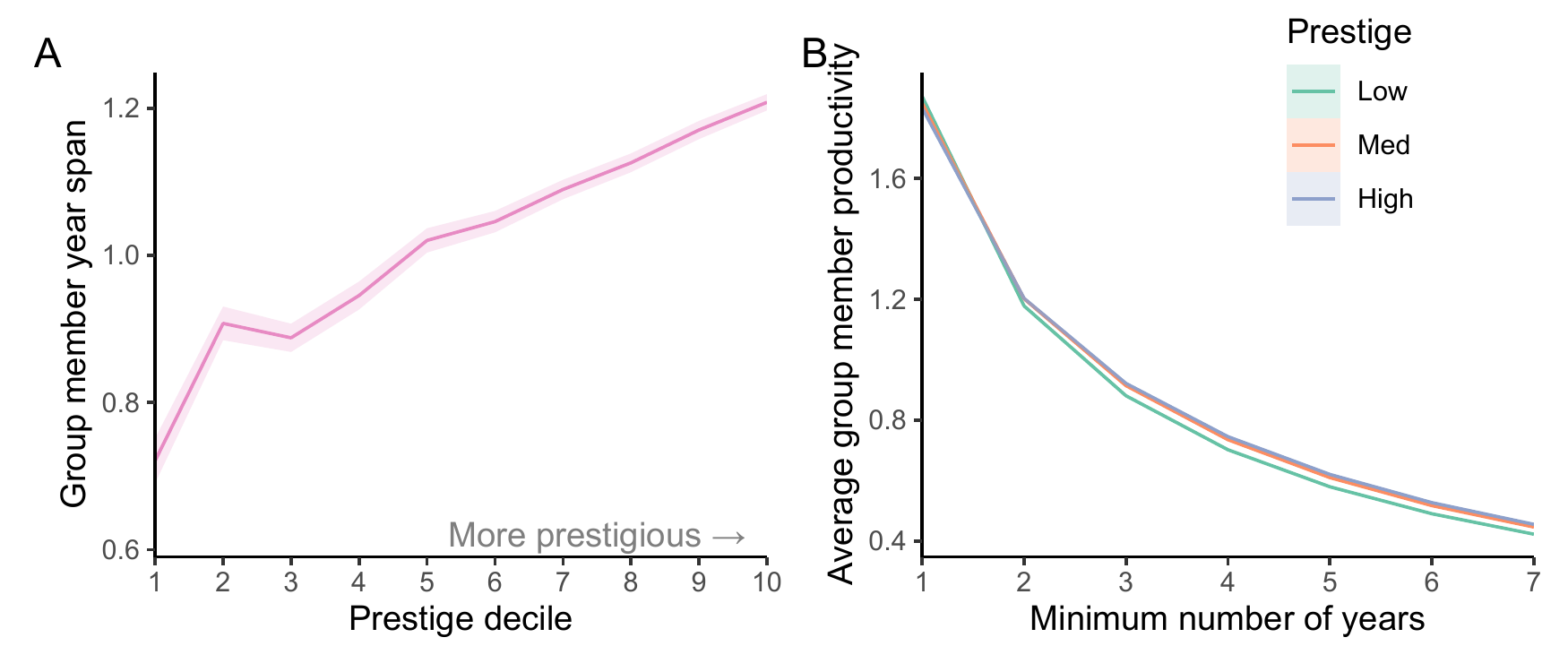}
    \caption{Non-faculty group member productivity with different assumed career lengths.
    (A) 
        Non-faculty group members at more prestigious institutions tend to have a higher productive range in the data.
        Displayed are means with 95\% confidence interval at each prestige decile. 
    (B) 
        When non-faculty have only one or two publications in the data in a short length of time, then we need to choose what we mean by their ``average" productivity.
        In particular, we can impose a minimum length of time that we consider a group member to be present, filling in years outside of that range as zeros.
        As we increase the minimum span of time, the average productivity declines, but nonlinearly, since some, but not all, group members already appear for at least that length of time.
        We find that increasing the minimum span helps distinguish between the productivity of more prestigious and less prestigious group members.
        Displayed are means with 95\% confidence intervals around the means for each minimum year span. 
    }
	\label{fig:supp_nonfac_min_years}
\end{figure*}

Group members at more elite institutions tend to exhibit a slightly higher average year span than group members at less prestigious institutions, with each increase in prestige decile associated with an average increase of $0.043$ years of year span (t-test, $p<0.001$; Fig.~\ref{fig:supp_nonfac_min_years}A).
This has implications for how we compute the average productivity of non-faculty in the data.
To compute the average productivity for each non-faculty group member, we divide the total number of their publications by some range of time.
If we use the empirical year span as the range of time, then non-faculty group members at less elite institutions will have a smaller average denominator in their productivity, artificially increasing their productivity.
For this reason, instead of using the year span directly, we use the minimum of the year span and some range of time, namely five years in Fig.~\ref{fig:decomp}A.
To check that our results are insensitive to the choice of the minimum year span, we plot the average productivities of different prestige strata by the minimum year span (Fig.~\ref{fig:supp_nonfac_min_years}B).

\begin{figure}[ht]
    \centering
    \includegraphics[width=0.3\linewidth]{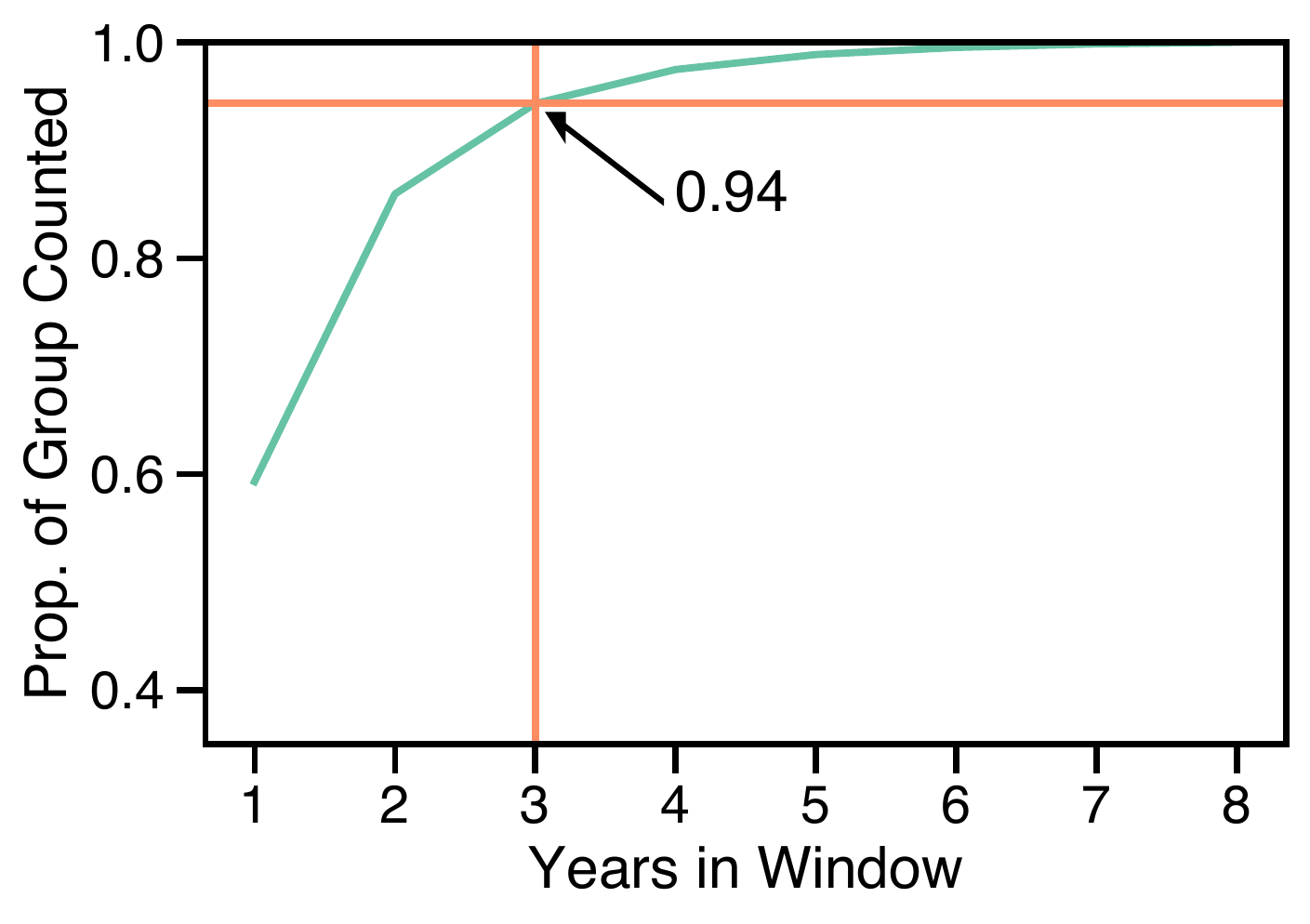}
	\caption{Choosing a window size for estimating group sizes. 
    Certain research group members may only publish with faculty every couple of years, and we would prefer to consider those people to be active group members during their inactivity, if they would later publish with the faculty again within a maximum length of time, which we set to be 7 years. Choosing a window size of 3 years includes 94\% of all active group members, while moving to four years increases the coverage to 97.6\%.}
    \label{fig:window_size}
\end{figure}

For the matching, we derive a measure for the group size for each faculty, based on these potential group members.
To reduce the extent to which our measure of group size is confounded by productivity, we window our measure of group size over several years; that is, we count the number of unique same-department non-faculty collaborators over a sliding window of $k$ years.
For example, certain research group members may not publish every year with faculty, but have gaps in their publication records.
To select a useful window, we consider the tradeoffs between window length and coverage to pick the smallest acceptable window (Fig.~\ref{fig:window_size}).
A window size of 3 only omits about 6\% of all groups, which we view as an acceptable range of error for this analysis.
To perform the actual construction, we slide this window of 3 or 4 years over the publication data, counting the number of unique group member collaborators of each faculty member in our dataset.

\subsection{Inferring genders from names}%
\label{sub:inferring_genders_from_names}

We inferred a gender (man, woman, or unknown) for each faculty member using first and last names.
First, we checked complete names against two offline dictionaries: a hand-annotated list of faculty employed at Business, Computer Science, and History departments~\cite{clauset2015systematic}, and the open-source python package gender-guesser~\cite{genderguesser2016}.
These dictionaries assigned each full name as either female, male, or unable to classify.
Second, where the dictionaries disagreed or where either dictionary was unable to assign a gender to the name, we queried Ethnea~\cite{torvik2016ethnea} and used the gender they assigned the name.
Using this approach we were able to assign genders to 88.3\% of faculty.
Faculty whose names could not be associated with a gender were excluded from the individual-level regressions but still included in other analyses.
We recognize that gender is nonbinary, although this procedure assigns binary (woman/man) labels to faculty.
This is a compromise between the technical limitations of name-based gender inference and the importance of studying gender inequality in science, and it is not intended to reinforce the gender binary.

\subsection{Counting cumulative productivity and cumulative group size}%
\label{sub:generating_productivity_group_size_curves}

We generate cumulative counts of productivity and group size in order to plot the relationship between the two across prestige strata (Fig.~\ref{fig:middle_links}C).
To start, we need to pick a reference point for starting the count $t_0$, which we take as the latter of either their degree year or the first year that someone appears in Web of Science with the same affiliation as the record we have in the employment data.
Then, to compute cumulative (group) productivity for a given year $t > t_0$, we add up the number of publications they have starting from the first year they appear in the data.
To count cumulative group size, we maintain a record of all of the past group members (same-department non-faculty collaborators) in a hash set data structure that have been seen up to year $t$, and we count the number of unique elements in that set.
We assume our publication data is complete in the sense that in years where we have no data for faculty in Web of Science, we assume they publish zero papers, and introduce zero new group members.

\subsection{Disciplines with and without collaboration norms}%
\label{sec:disciplines_with_and_without_collaboration_norms}

The labor advantage mechanism depends on graduate, postdoctoral, and other non-faculty labor contributing to faculty productivity.
Norms differ across academic disciplines on how closely faculty collaborate with research group members on projects, and even among disciplines where such collaborations do occur, on whether faculty tend to share credit with students in the form of coauthorships.
We have relied on sociological evidence to separate disciplines into these two categories in Table~\ref{tab:discipline_norms}, with broadly social sciences, humanities, and mathematics as the disciplines without research group collaboration norms, and the natural and medical sciences and engineering as the disciplines with collaboration norms~\cite{lariviere2012shoulders, maher2013factors, kamler2008rethinking, shrum2007structures, hagstrom1965scientific, hargens1975patterns, kyvik1994teaching}.


The concept of an academic discipline has been continuously redefined by practitioners~\cite{sugimoto2015kaleidoscope}, and our use of the word ``discipline" may be closer to what others may call a ``disciplinary grouping", where we combine, for example, computer and information sciences into ``computational sciences".
To test the sensitivity of the results to our choice of aggregation, we perform a separate bipartition at the more detailed taxonomic level of a ``field", which we use as a term to be more granular than ``discipline"~\cite{fry2006scholarly, carayol2005academic}.
The results were robust to this finer partitioning scheme, and the main results in the paper were given in terms of the broader partitions.
In the field bipartition, all of engineering and natural sciences remained as having collaboration norms, but we identified as fields with collaboration norms: General Psychology, Neuroscience,
Computer Science, Statistics, Biostatistics, Information Science, Finance, Animal Science, Food Science, Agronomy, Soil Science, Nursing, Veterinary Medical Sciences, Physiology, Pharmacy, Epidemiology, Communication Disorders and Sciences, Nutrition Sciences, and Pharmacology.
The remaining fields without collaboration norms were then Mathematics, Political Science, General Economics, Sociology, Anthropology, Geography, and Agricultural Economics.

\section{Modeling details}
\subsection{Matching}
\label{supptext}

To perform our matching analysis, we first construct a dataset of mid-career moves.
First, we expand our consideration to include 25 disciplines, rather than only the 17 in the NSF GSS, to observe as many mid-career moves as possible. 
However, this eliminates our ability to use available departmental labor as a matching variable.
Thus, we use prestige as a proxy for labor availability.
Since our dataset consists of a longitudinal census of faculty between the years 2011 and 2017, we can look for people who changed institutions during that time.
We avoid looking at cases where someone only changed departments within the same institution, since that may not be a sufficiently large change in terms of working conditions and available resources.
We require that faculty only move once during the period of observation, that they be present in at least four years of employment data, and that they stay within the same academic discipline before and after their move.
This leaves 5,709 mid-career moves.
However, not all of these people have publication records available in WoS for the years before and after their move.
Once we impose the requirement that people have publications within the four years before their move, as well as two years after their move, we are left with 2,316 people.

The ideal experiment to assess the impact of available labor on productivity would be to randomly allocate funded graduate and postdoctoral labor to departments.
Since that experiment is difficult to run, we consider an observational study based on faculty mobility instead: we use mid-career moves in our dataset as a quasi-random allocation of faculty into high and low labor availability environments.
However, we only have departmental labor data for the disciplines that are included in the NSF survey, and we have few enough mid-career moves that we prefer to use as much data as possible.
To that end, we rely on prestige as a proxy for labor availability in the matching (see Supplementary Material), and the treatment condition consists of faculty who move upwards in the prestige hierarchy, and the control condition are faculty who move to a less prestigious institution.
The outcome variable is the productivity with group members, which indirectly measures the labor advantage effect, since group members themselves are not more productive at elite institutions.
In the matching, we control for academic discipline, rank (assistant, associate, or full professor), pre-move prestige, pre-move productivity, and pre-move group size.
Then, we look at with and without group productivity two years after the move.
Picking two years allows us to avoid the noise associated with the move itself, while leaving enough people in to make a precise estimate.

\begin{figure}[ht]
    \centering
    \includegraphics[width=0.5\linewidth]{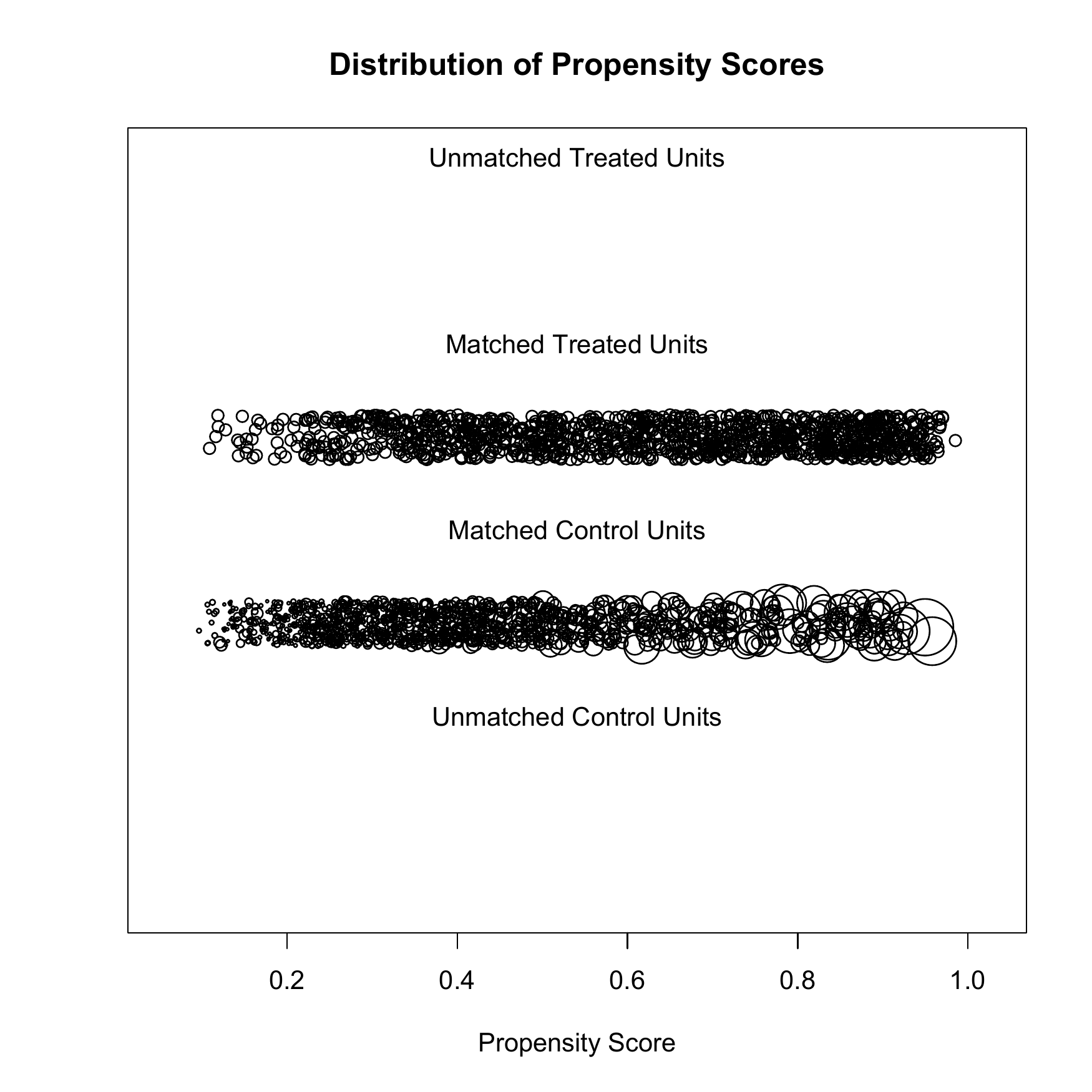}
    \caption{Distribution of propensity scores.
    The treatment and control groups exhibit a similar distribution of propensity scores after the full propensity score matching.}
    \label{fig:matching_jitter}
\end{figure}

\begin{figure}[ht]
    \centering
    \includegraphics[width=0.8\linewidth]{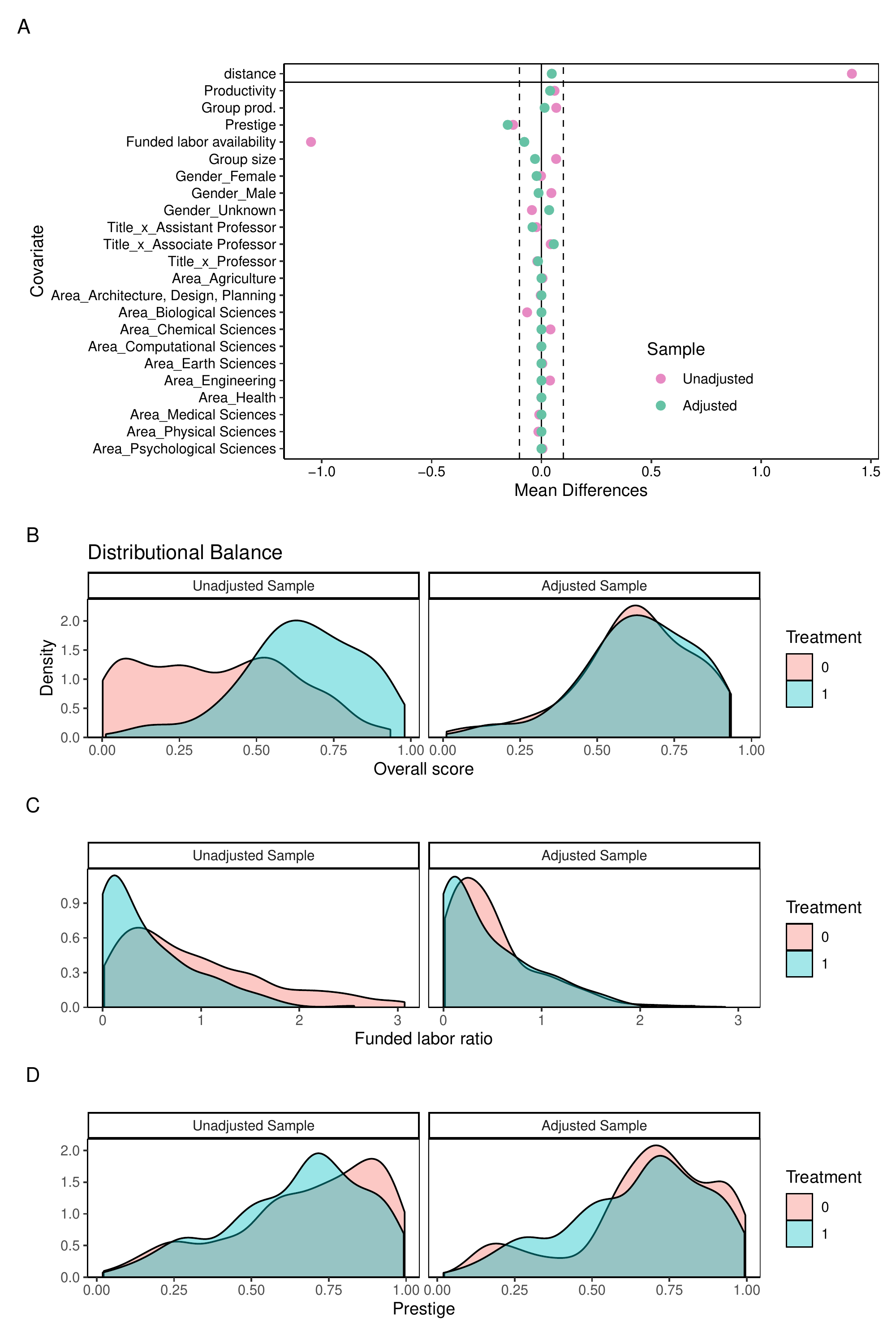}
    \caption{Balance of covariates before and after matching.
        (A) Diagnostic test of covariate balance before and after matching, as well as overall propensity score difference between groups (``difference" in top row).
        Prior to matching, we had substantial imbalances in both covariates and overall distance (pink), that were largely corrected to an acceptable level of below 0.1 mean difference after adjustment (green).
        The exception is that prestige remained imbalanced after matching, which we diagnose further below.
        (B) Overall distributional balance of propensity scores is corrected by the matching.
        (C) The most imbalanced pre-adjustment variable, the logarithm of the ratio of funded labor to faculty in departments, is adequately balanced by the matching.
        (D) The only variable that remained imbalanced after adjustments was prestige.
        In particular, upward movers are more likely to come from less prestigious institutions, and downward movers are more likely to come from more prestigious institutions, due to the inability of those at the bottom of the prestige hierarchy to move downward and the inability of those at the top of the prestige hierarchy to move upward.
    }
    \label{fig:matching_balance}
\end{figure}

We first attempted 1:1 nearest neighbor propensity score matching, but found inadequate balance.
Instead, we used full matching on the propensity score, which yielded adequate balance.
We find similar distributions of propensity scores between the matched control and treated units (Fig.~\ref{fig:matching_jitter}).
Prior to matching, we had substantial imbalance in the pre-move prestige between cohorts, and slight imbalance in total and group productivity. 
The difference in the propensity score (``distance") between groups before matching was also substantial.
After matching, imbalance in all covariates and the propensity score is reduced to an acceptable threshold of below 0.1 mean difference. (Fig.~\ref{fig:matching_balance}).

To estimate the treatment effect of prestige on research group size, we used a linear regression on the number of papers published with department non-faculty collaborators as the outcome, a binary indicator variable whether the place that a faculty moved to was higher prestige as the treatment, and including covariates from before the move such as productivity, productivity with group members, prestige, research discipline, unique same-department non-faculty collaborators on papers, gender, and title.
We include the full matching weights in the estimation.
We estimate standard errors from a cluster-robust variance using the matching stratum membership as the clustering variable, and perform a t-test on whether the coefficient of the treatment variable is statistically significantly different from zero.

As a robustness test, we also considered the impact on productivity three, instead of two, years after the move.
Since this requires an additional year of data, our sample size decreased from 2,316 to 805, with 417 (51.8\%) moves up the prestige hierarchy.
We followed the exact same matching procedure as before, and we found adequate balance due to the full propensity score adjustment.
Similar to the two year case, we found an effect on group productivity of moving up rather than down the prestige hierarchy, but the signal became weaker, with upward movers publishing $0.498$ more papers on average three years after moving than downward movers (t-test, $p=0.00062$). 
On the other hand, an effect on individual productivity appeared, with individual productivity for upward movers exceeding downward movers by $0.471$ on average on the third year after the move (t-test, $p<0.001$). 
Further investigation would be necessary to illuminate whether this is due to noise, or whether it reflects faculty choosing to prioritize the ramping up of a new research group after moving before working on individual publications.

We found that faculty tended to have smaller groups after they move than before they move in general, reflecting a general decrease or ``shock" to group size immediately after making a mid-career move.
This suggested that matching faculty who make mid-career moves to control faculty who do not move would not be useful, since we would likely find decreases in both conditions.

\subsection{Alternate model specifications}

An ideal dataset for analyzing the impact of labor on productivity would be one that reported the research group size of each faculty member for each year.
However, such a dataset would be difficult to define and construct, since the exact dates that students or postdocs begin working with a faculty member are not necessarily clear:
an informal advising relationship may progress into a formal one after a trial period, for example through a series of laboratory rotations during the beginning of a graduate program, or from the student attending the class of a faculty and being advised on a class project.
From a data gathering perspective, faculty often only report the graduation year of their PhD students on their CVs, reflecting and compounding on the difficulty of pinning down the start of these formal relationships.
On the other hand, aggregated departmental data on labor availability can be reliably sourced from our census of tenure-track faculty and NSF survey data for certain disciplines.
We thus perform an analysis at the aggregated department level, where we use a regression to consider the relative impact of departmental variables such as the ratio of funded (and unfunded) researchers to faculty within the department on departmental productivity.

\begin{figure*}[ht]
    \centering
    \includegraphics[width=0.8\linewidth]{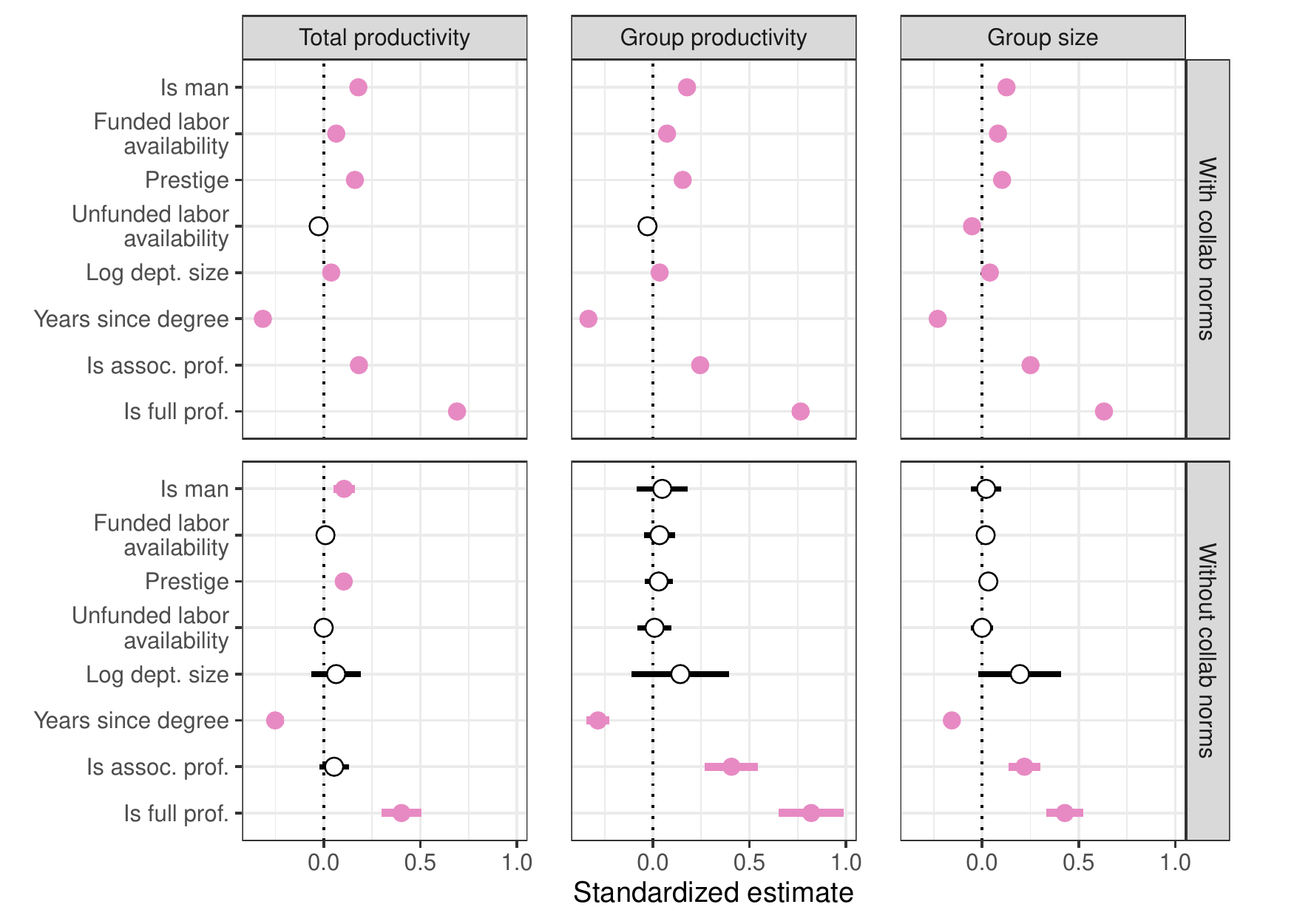}
    \caption{Coefficients of individual-level regression.
        Performing a Poisson regression at the individual faculty level controlling for discipline using fixed effects and clustering standard errors within departments, we report the coefficients of standardized (zero mean and unit variance) individual, departmental, and institutional covariates in predicting departmental productivity, group productivity, and group sizes, in disciplines with and without collaboration norms, with 95\% confidence intervals.
        Statistically significant coefficients at $p<0.05$ are depicted in pink with a filled-in circle.
        The availability of funded labor has a significant impact on all dependent variables, even after controlling for prestige, especially in disciplines with collaboration norms.
    }
	\label{fig:supp_coefs_ppl}
\end{figure*}

\begin{figure*}[ht]
    \centering
    \includegraphics[width=0.8\linewidth]{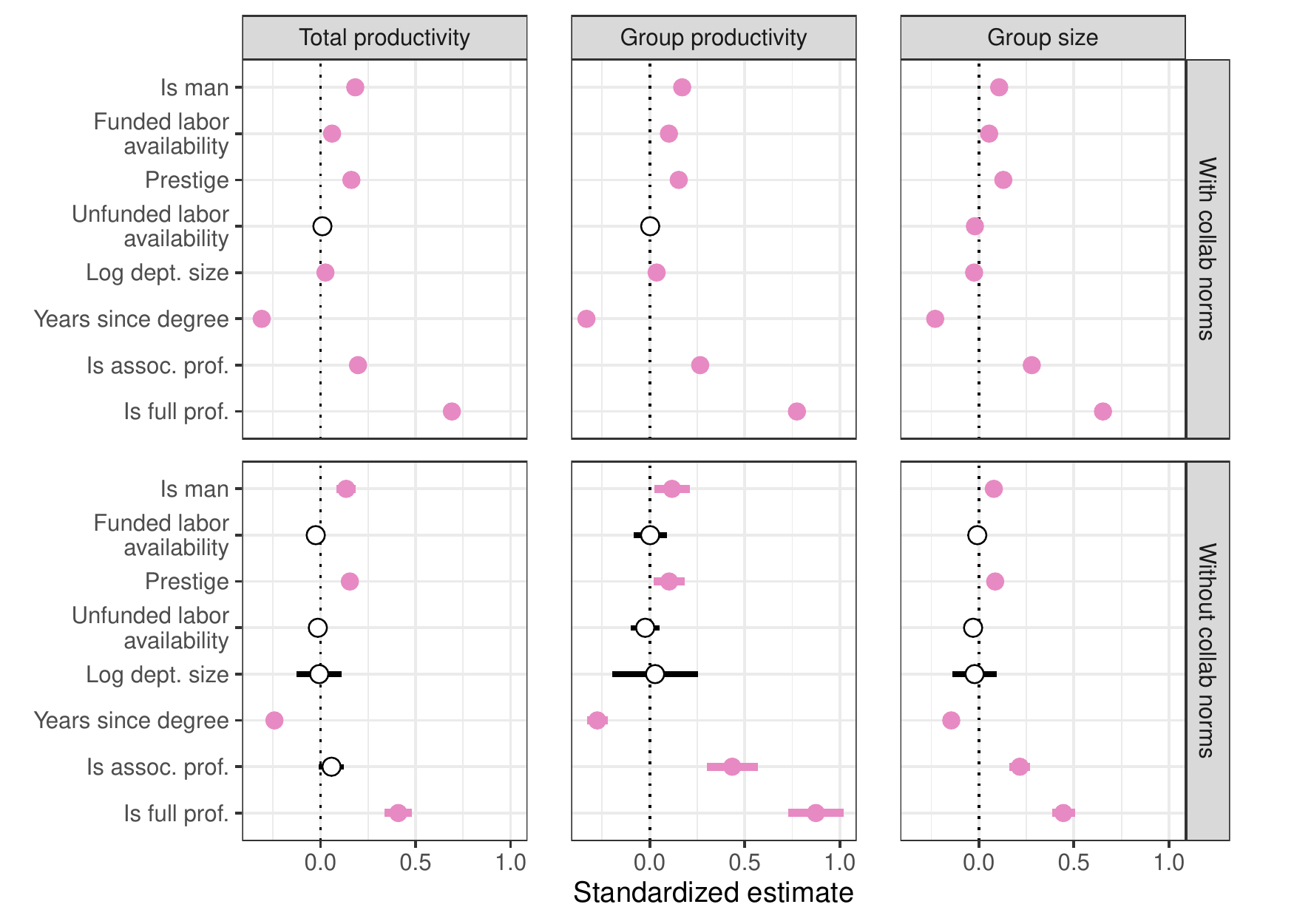}
    \caption{Coefficients of hierarchical regression.
        Performing a hierarchical regression using individual faculty as our unit, we report the coefficients of standardized (zero mean and unit variance) individual, departmental, and institutional covariates in predicting departmental productivity, group productivity, and group sizes, in disciplines with and without collaboration norms, with 95\% confidence intervals.
        Statistically significant coefficients at $p<0.05$ are depicted in pink with a filled-in circle.
        The availability of funded labor has a significant impact on all dependent variables, even after controlling for prestige, especially in disciplines with collaboration norms.
    }
	\label{fig:supp_coefs_ppl_hierc}
\end{figure*}

To test the robustness of our regression specification, we run alternate models using individual-level data.
Associating individuals with additional metadata relevant to their productivity, such as their last known rank, their gender inferred from their name, and the number of years since their degree, we average their productivity to create a dataset for predicting individual average productivity from both individual and institutional characteristics.
This produces three nested levels: individual, institution-discipline, and discipline.
A simple model with few assumptions is a Poisson regression where disciplines are controlled using fixed effects, and then we adjust the standard errors to account for clustering at the institution-discipline level (Table~\ref{tab:ppl_pois_reg}, Fig.~\ref{fig:supp_coefs_ppl}).
Another natural choice, though with more assumptions, is to model the hierarchy explicitly through a generalized hierarchical linear model, where institution-disciplines and disciplines are the two levels (Table~\ref{tab:hierc_pois_reg}, Fig.~\ref{fig:supp_coefs_ppl_hierc}).
In both models, we find that funded labor is statistically and practically significant in explaining the dependent variables, although only in disciplines with collaboration norms.

\begin{figure*}[ht]
    \centering
    \includegraphics[width=0.7\linewidth]{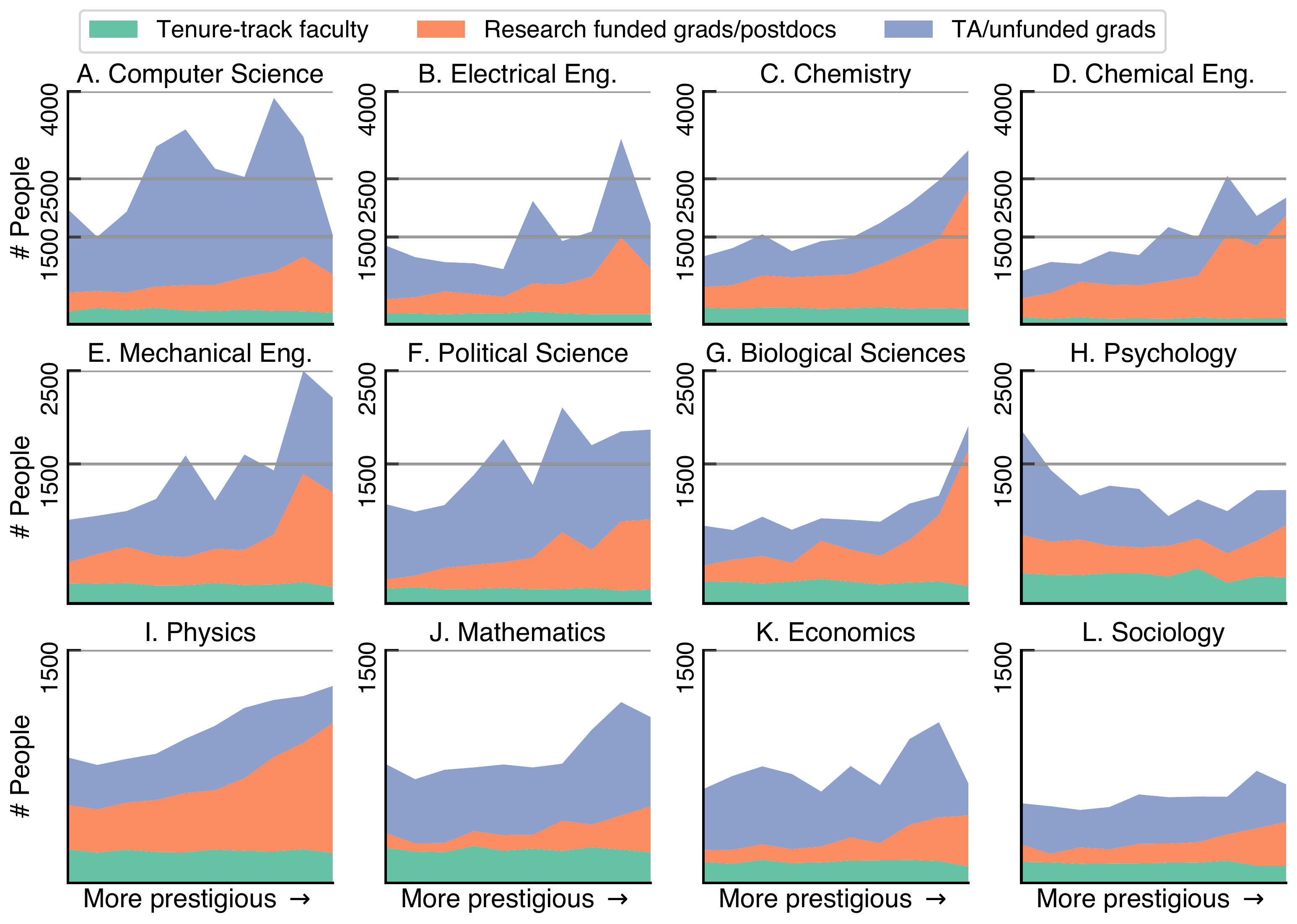}
    \caption{Labor advantages exist in absolute terms across research fields. We group departments within each field into prestige deciles with the goal of keeping the counts of tenure-track faculty in each decile roughly the same, to highlight the ratio of funded and unfunded labor to faculty.
        Across most fields, the ratio of funded graduate students and postdoctoral researchers (orange) to faculty increases steadily with prestige, while graduate teaching assistants and self-funded graduate students (purple) remains relatively constant with prestige.
    }
	\label{fig:supp_absolute_counts}
\end{figure*}

\begin{figure}[ht]
    \centering
    \includegraphics[width=0.5\linewidth]{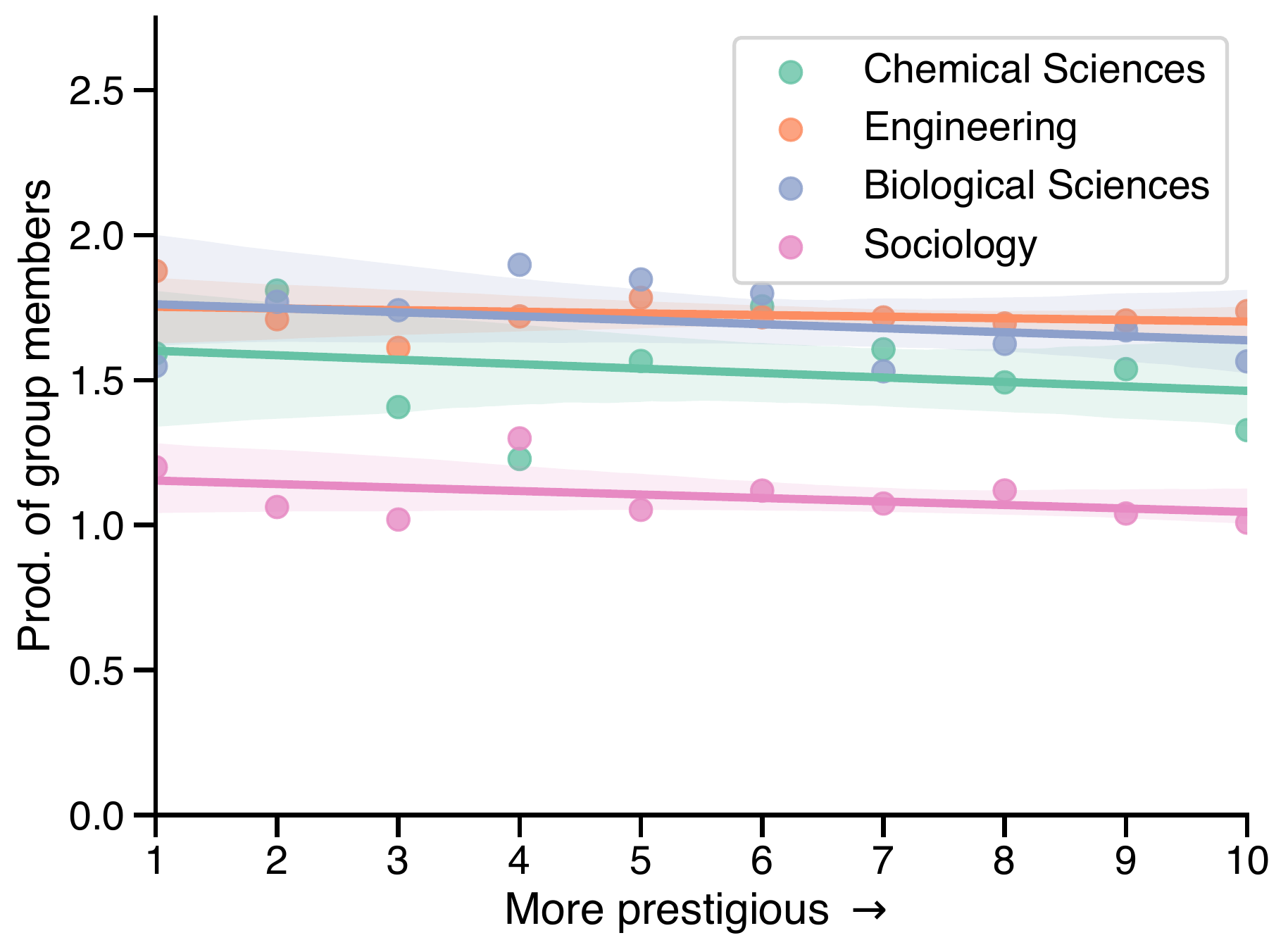}
    \caption{Non-faculty are roughly equally productive across prestige, across disciplines.
        Disaggregating non-faculty into four disciplines, using no minimum year span, we find that non-faculty collaborators of faculty do not publish more papers in collaboration with faculty in their department at more elite institutions.
        Displayed are average productivities for each prestige decile, and the 95\% confidence interval for the ordinary linear regression through the ten points for each discipline.
    }
	\label{fig:supp_decomp}
\end{figure}

\begin{table}[]
    \centering
\begin{tabular}{ll}
\hline
\multicolumn{1}{|l|}{\textbf{\begin{tabular}[c]{@{}l@{}}Disciplines with\\ collaboration norms\end{tabular}}} & \multicolumn{1}{l|}{\textbf{\begin{tabular}[c]{@{}l@{}}Disciplines without\\ collaboration norms\end{tabular}}} \\ \hline
\multicolumn{1}{|l|}{Biological Sciences}    & \multicolumn{1}{l|}{Economics}                     \\ \hline
\multicolumn{1}{|l|}{Engineering}            & \multicolumn{1}{l|}{Mathematical Sciences}         \\ \hline
\multicolumn{1}{|l|}{Medical Sciences}       & \multicolumn{1}{l|}{Anthropology} \\ \hline
\multicolumn{1}{|l|}{Psychological Sciences} & \multicolumn{1}{l|}{Political Science}             \\ \hline
\multicolumn{1}{|l|}{Physical Sciences}      & \multicolumn{1}{l|}{Sociology}                       \\ \hline
\multicolumn{1}{|l|}{Chemical Sciences}      & \multicolumn{1}{l|}{Geography}                  \\ \hline
\multicolumn{1}{|l|}{Computational Science}  & \multicolumn{1}{l|}{}                     \\ \hline
\multicolumn{1}{|l|}{Health}                 & \multicolumn{1}{l|}{}                                          \\ \hline
\multicolumn{1}{|l|}{Agriculture}               & \multicolumn{1}{l|}{}                                        \\ \hline 
\multicolumn{1}{|l|}{Earth Sciences}         & \multicolumn{1}{l|}{}                                               \\ \hline
\multicolumn{1}{|l|}{\begin{tabular}[c]{@{}l@{}}Architecture, Design,\\Planning\end{tabular}}   & \multicolumn{1}{l|}{}                                          \\ \hline
\end{tabular}

\caption{Disciplines by collaboration norms. Disciplines are separated into those where research groups tend to collaborate with principal investigators on projects, contributing to the PI's productivity (left) and those where those expectations aren't widely held (right).}
\label{tab:discipline_norms}
\end{table}

\begin{table*}
\begin{center}
    
    \begin{tabular}{l c c c c c c}
\hline
 & (p/c) & (gp/c) & (gs/c) & (p/nc) & (gp/nc) & (gs/nc) \\
\hline
Is private           & $-0.06$      & $-0.05$      & $-0.07$      & $-0.10^{*}$ & $-0.50^{***}$ & $-0.19^{***}$ \\
                     & $(0.10)$     & $(0.09)$     & $(0.07)$     & $(0.05)$    & $(0.11)$      & $(0.03)$      \\
Num. faculty         & $0.08$       & $0.12^{*}$   & $0.10^{**}$  & $-0.06$     & $-0.16$       & $-0.04$       \\
                     & $(0.06)$     & $(0.06)$     & $(0.04)$     & $(0.04)$    & $(0.10)$      & $(0.04)$      \\
Prestige             & $0.06$       & $0.04$       & $0.07^{**}$  & $0.07$      & $0.05$        & $-0.03$       \\
                     & $(0.06)$     & $(0.05)$     & $(0.02)$     & $(0.05)$    & $(0.06)$      & $(0.05)$      \\
Unfunded labor ratio & $-0.05$      & $-0.01$      & $-0.00$      & $-0.02$     & $-0.11$       & $-0.04$       \\
                     & $(0.07)$     & $(0.07)$     & $(0.05)$     & $(0.03)$    & $(0.07)$      & $(0.03)$      \\
Funded labor ratio   & $0.15^{***}$ & $0.17^{***}$ & $0.13^{***}$ & $0.02$      & $0.06$        & $0.09^{*}$    \\
                     & $(0.03)$     & $(0.03)$     & $(0.02)$     & $(0.04)$    & $(0.09)$      & $(0.04)$      \\
\hline
AIC                  & $1380.38$    & $1123.26$    & $2149.41$    & $636.35$    & $321.89$      & $757.90$      \\
BIC                  & $1438.13$    & $1181.01$    & $2207.15$    & $676.41$    & $361.95$      & $797.96$      \\
Log Likelihood       & $-676.19$    & $-547.63$    & $-1060.70$   & $-307.17$   & $-149.95$     & $-367.95$     \\
\hline
\multicolumn{7}{l}{\scriptsize{$^{***}p<0.001$; $^{**}p<0.01$; $^{*}p<0.05$}}
\end{tabular}

\caption{Poisson regression in departments separated by dependent variable (p=productivity, gp=group productivity, gs=group size) and the partition of the disciplines (c=disciplines with collaboration norms, nc=disciplines with no collaboration norms), using only one-to-one matches between employment data and NSF GSS (n=739).
The funded labor covariate is the base 2 logarithm of the ratio of funded researchers (including tenure-track faculty) to the tenure-track faculty.
The unfunded labor covariate is the base 2 logarithm of the ratio of unfunded researchers to the tenure-track faculty.
All continuous variables are standardized to have zero mean and unit variance.
}
\label{tab:reg}
\end{center}
\end{table*}

\begin{table*}
\begin{center}
\begin{tabular}{l c c c c c c}
\hline
 & (p/c) & (gp/c) & (gs/c) & (p/nc) & (gp/nc) & (gs/nc) \\
\hline
Is private           & $-0.05^{***}$ & $-0.06^{*}$  & $-0.04$       & $-0.03$    & $-0.21$    & $-0.12^{***}$ \\
                     & $(0.01)$      & $(0.02)$     & $(0.02)$      & $(0.03)$   & $(0.15)$   & $(0.03)$      \\
Num. faculty         & $0.07^{***}$  & $0.08^{***}$ & $0.09^{***}$  & $0.04$     & $-0.02$    & $0.11$        \\
                     & $(0.02)$      & $(0.02)$     & $(0.01)$      & $(0.08)$   & $(0.16)$   & $(0.10)$      \\
Prestige             & $0.14^{***}$  & $0.14^{***}$ & $0.09^{***}$  & $0.07^{*}$ & $0.00$     & $0.02$        \\
                     & $(0.03)$      & $(0.03)$     & $(0.02)$      & $(0.03)$   & $(0.04)$   & $(0.03)$      \\
Unfunded labor ratio & $-0.06^{***}$ & $-0.06^{**}$ & $-0.07^{***}$ & $-0.01$    & $-0.02$    & $-0.01$       \\
                     & $(0.02)$      & $(0.02)$     & $(0.02)$      & $(0.03)$   & $(0.04)$   & $(0.03)$      \\
Funded labor ratio   & $0.12^{***}$  & $0.14^{***}$ & $0.13^{***}$  & $0.06$     & $0.12^{*}$ & $0.08^{***}$  \\
                     & $(0.02)$      & $(0.02)$     & $(0.01)$      & $(0.04)$   & $(0.05)$   & $(0.02)$      \\
\hline
AIC                  & $3505.15$     & $2784.29$    & $5413.71$     & $1080.54$  & $550.30$   & $1352.12$     \\
BIC                  & $3587.79$     & $2866.92$    & $5496.35$     & $1127.06$  & $596.82$   & $1398.63$     \\
Log Likelihood       & $-1736.58$    & $-1376.14$   & $-2690.86$    & $-529.27$  & $-264.15$  & $-665.06$     \\
\hline
\multicolumn{7}{l}{\scriptsize{$^{***}p<0.001$; $^{**}p<0.01$; $^{*}p<0.05$}}
\end{tabular}
\caption{Poisson regression in institution-disciplines separated by dependent variable (p=productivity, gp=group productivity, gs=group size) and the partition of the disciplines (c=disciplines with collaboration norms, nc=disciplines with no collaboration norms), allowing many-to-many matches between employment data and NSF GSS (n=1800).
The funded labor covariate is the base 2 logarithm of the ratio of funded researchers (including tenure-track faculty) to the tenure-track faculty.
The unfunded labor covariate is the base 2 logarithm of the ratio of unfunded researchers to the tenure-track faculty.
All continuous variables are standardized to have zero mean and unit variance.
}
\label{tab:reg_expansive}
\end{center}
\end{table*}

\begin{table*}
    \begin{center}
\begin{tabular}{l c c c c}
\hline
 & (first/c) & (last/c) & (first/nc) & (last/nc) \\
\hline
Is private           & $-0.05$   & $-0.16^{*}$  & $0.07$    & $-0.00$      \\
                     & $(0.15)$  & $(0.07)$     & $(0.05)$  & $(0.07)$     \\
Num. faculty         & $0.01$    & $0.07$       & $-0.05$   & $-0.03$      \\
                     & $(0.07)$  & $(0.05)$     & $(0.05)$  & $(0.05)$     \\
Prestige             & $0.02$    & $0.17^{***}$ & $0.04$    & $0.09^{***}$ \\
                     & $(0.07)$  & $(0.04)$     & $(0.05)$  & $(0.03)$     \\
Unfunded labor ratio & $-0.09$   & $-0.01$      & $0.01$    & $-0.03$      \\
                     & $(0.07)$  & $(0.06)$     & $(0.03)$  & $(0.02)$     \\
Funded labor ratio   & $0.01$    & $0.10^{**}$  & $-0.03$   & $0.00$       \\
                     & $(0.04)$  & $(0.03)$     & $(0.04)$  & $(0.04)$     \\
\hline
AIC                  & $477.01$  & $844.79$     & $430.92$  & $413.39$     \\
BIC                  & $534.75$  & $902.54$     & $470.98$  & $453.45$     \\
Log Likelihood       & $-224.50$ & $-408.40$    & $-204.46$ & $-195.70$    \\
\hline
\multicolumn{5}{l}{\scriptsize{$^{***}p<0.001$; $^{**}p<0.01$; $^{*}p<0.05$}}
\end{tabular}
\caption{Poisson regression in departments separated by dependent variable (first=annual papers authored as first author, last=annual papers authored as last author) and the partition of the disciplines (c=disciplines with collaboration norms, nc=disciplines with no collaboration norms).
The funded labor covariate is the base 2 logarithm of the ratio of funded researchers (including tenure-track faculty) to the tenure-track faculty.
The unfunded labor covariate is the base 2 logarithm of the ratio of unfunded researchers to the tenure-track faculty.
All continuous variables are standardized to have zero mean and unit variance.
}
\label{tab:reg_firstlast}
\end{center}
\end{table*}

\begin{table*}
\begin{center}
    
    \begin{tabular}{l c c c c c c c c}
\hline
 & (gs/c) & (gs/c) & (gs/c) & (gs/nc) & (gp/c) & (gp/nc) & (p/c) & (p/nc) \\
\hline
Is full prof.               & $0.64^{***}$  & $0.63^{***}$  & $0.63^{***}$  & $0.43^{***}$  & $0.76^{***}$  & $0.82^{***}$  & $0.69^{***}$  & $0.40^{***}$  \\
                            & $(0.02)$      & $(0.02)$      & $(0.02)$      & $(0.05)$      & $(0.02)$      & $(0.09)$      & $(0.02)$      & $(0.05)$      \\
Is assoc. prof.             & $0.25^{***}$  & $0.25^{***}$  & $0.25^{***}$  & $0.22^{***}$  & $0.24^{***}$  & $0.41^{***}$  & $0.18^{***}$  & $0.05$        \\
                            & $(0.02)$      & $(0.02)$      & $(0.01)$      & $(0.04)$      & $(0.02)$      & $(0.07)$      & $(0.02)$      & $(0.04)$      \\
Years since degree          & $-0.23^{***}$ & $-0.23^{***}$ & $-0.23^{***}$ & $-0.16^{***}$ & $-0.33^{***}$ & $-0.28^{***}$ & $-0.31^{***}$ & $-0.25^{***}$ \\
                            & $(0.01)$      & $(0.01)$      & $(0.01)$      & $(0.02)$      & $(0.01)$      & $(0.03)$      & $(0.01)$      & $(0.02)$      \\
Log dept. size              & $0.07^{***}$  & $0.02$        & $0.04^{**}$   & $0.20$        & $0.03^{*}$    & $0.14$        & $0.04^{**}$   & $0.06$        \\
                            & $(0.01)$      & $(0.01)$      & $(0.01)$      & $(0.11)$      & $(0.01)$      & $(0.13)$      & $(0.01)$      & $(0.06)$      \\
Unfunded labor
availability & $-0.07^{***}$ & $-0.01$       & $-0.05^{**}$  & $-0.00$       & $-0.03$       & $0.01$        & $-0.03$       & $0.00$        \\
                            & $(0.02)$      & $(0.02)$      & $(0.02)$      & $(0.03)$      & $(0.02)$      & $(0.04)$      & $(0.02)$      & $(0.02)$      \\
Prestige                    &               & $0.13^{***}$  & $0.10^{***}$  & $0.03$        & $0.15^{***}$  & $0.03$        & $0.16^{***}$  & $0.10^{***}$  \\
                            &               & $(0.01)$      & $(0.01)$      & $(0.02)$      & $(0.01)$      & $(0.04)$      & $(0.01)$      & $(0.02)$      \\
Funded labor
availability   & $0.14^{***}$  &               & $0.08^{***}$  & $0.02$        & $0.07^{***}$  & $0.03$        & $0.06^{***}$  & $0.01$        \\
                            & $(0.01)$      &               & $(0.01)$      & $(0.02)$      & $(0.02)$      & $(0.04)$      & $(0.02)$      & $(0.02)$      \\
Is man                      & $0.12^{***}$  & $0.13^{***}$  & $0.13^{***}$  & $0.02$        & $0.18^{***}$  & $0.05$        & $0.18^{***}$  & $0.11^{***}$  \\
                            & $(0.01)$      & $(0.01)$      & $(0.01)$      & $(0.04)$      & $(0.02)$      & $(0.07)$      & $(0.01)$      & $(0.03)$      \\
\hline
AIC                         & $378641.65$   & $377382.39$   & $376591.83$   & $30716.39$    & $152875.57$   & $11674.50$    & $197346.55$   & $24045.61$    \\
BIC                         & $378800.75$   & $377541.50$   & $376759.77$   & $30815.43$    & $153043.52$   & $11773.55$    & $197514.50$   & $24144.65$    \\
Log Likelihood              & $-189302.82$  & $-188673.19$  & $-188276.91$  & $-15344.19$   & $-76418.79$   & $-5823.25$    & $-98654.28$   & $-12008.80$   \\
\hline
\multicolumn{9}{l}{\scriptsize{$^{***}p<0.001$; $^{**}p<0.01$; $^{*}p<0.05$}}
\end{tabular}

\caption{As an alternate specification to the institution-discipline regressions, here we present the regression coefficients from a regression on individual faculty.
        We use a Poisson regression at the individual faculty level controlling for discipline using fixed effects and clustering standard errors within departments, we report the coefficients of standardized (zero mean and unit variance) individual, departmental, and institutional covariates in predicting departmental productivity, group productivity, and group sizes, in disciplines with and without collaboration norms, with 95\% confidence intervals.
    The model names correspond to the dependent variable (p=productivity, gp=group productivity, gs=group size) and the partition of the disciplines (c=disciplines with collaboration norms, nc=disciplines with no collaboration norms).
    We also include versions of the p/c model without the prestige and labor variables.
All continuous variables are standardized to have zero mean and unit variance.
}
\label{tab:ppl_pois_reg}
\end{center}
\end{table*}

\begin{table*}
\begin{center}

\begin{tabular}{l c c c c c c c c}
\hline
 & (gs/c) & (gs/c) & (gs/c) & (gs/nc) & (gp/c) & (gp/nc) & (p/c) & (p/nc) \\
\hline
Is full prof.                         & $0.65^{***}$  & $0.65^{***}$  & $0.65^{***}$  & $0.44^{***}$  & $0.59^{***}$  & $0.87^{***}$  & $0.69^{***}$  & $0.41^{***}$  \\
                                      & $(0.01)$      & $(0.01)$      & $(0.01)$      & $(0.03)$      & $(0.01)$      & $(0.07)$      & $(0.01)$      & $(0.04)$      \\
Is assoc. prof.                       & $0.28^{***}$  & $0.28^{***}$  & $0.28^{***}$  & $0.22^{***}$  & $0.21^{***}$  & $0.43^{***}$  & $0.20^{***}$  & $0.06$        \\
                                      & $(0.01)$      & $(0.01)$      & $(0.01)$      & $(0.03)$      & $(0.01)$      & $(0.07)$      & $(0.01)$      & $(0.03)$      \\
Years since degree                    & $-0.23^{***}$ & $-0.23^{***}$ & $-0.23^{***}$ & $-0.15^{***}$ & $-0.24^{***}$ & $-0.28^{***}$ & $-0.31^{***}$ & $-0.24^{***}$ \\
                                      & $(0.00)$      & $(0.00)$      & $(0.00)$      & $(0.01)$      & $(0.00)$      & $(0.03)$      & $(0.00)$      & $(0.01)$      \\
Log dept. size                        & $-0.01$       & $-0.04^{***}$ & $-0.03^{**}$  & $-0.02$       & $0.03^{*}$    & $0.03$        & $0.03^{**}$   & $-0.01$       \\
                                      & $(0.01)$      & $(0.01)$      & $(0.01)$      & $(0.06)$      & $(0.01)$      & $(0.12)$      & $(0.01)$      & $(0.06)$      \\
Unfunded labor
availability           & $-0.02^{**}$  & $-0.01$       & $-0.02^{**}$  & $-0.03$       & $0.01$        & $-0.02$       & $0.01$        & $-0.01$       \\
                                      & $(0.01)$      & $(0.01)$      & $(0.01)$      & $(0.02)$      & $(0.01)$      & $(0.04)$      & $(0.01)$      & $(0.02)$      \\
Prestige                              &               & $0.14^{***}$  & $0.13^{***}$  & $0.09^{***}$  & $0.14^{***}$  & $0.10^{*}$    & $0.16^{***}$  & $0.15^{***}$  \\
                                      &               & $(0.01)$      & $(0.01)$      & $(0.02)$      & $(0.01)$      & $(0.04)$      & $(0.01)$      & $(0.02)$      \\
Funded labor
availability             & $0.07^{***}$  &               & $0.05^{***}$  & $-0.01$       & $0.09^{***}$  & $0.00$        & $0.06^{***}$  & $-0.03$       \\
                                      & $(0.01)$      &               & $(0.01)$      & $(0.02)$      & $(0.01)$      & $(0.04)$      & $(0.01)$      & $(0.02)$      \\
Is man                                & $0.11^{***}$  & $0.11^{***}$  & $0.11^{***}$  & $0.08^{***}$  &               & $0.12^{*}$    & $0.18^{***}$  & $0.13^{***}$  \\
                                      & $(0.00)$      & $(0.00)$      & $(0.00)$      & $(0.02)$      &               & $(0.05)$      & $(0.01)$      & $(0.03)$      \\
\hline
Num. obs.                             & $50991$       & $50991$       & $50991$       & $8731$        & $67707$       & $8731$        & $50991$       & $8731$        \\
Num. groups: Department:Discipline     & $3750$        & $3750$        & $3750$        & $635$         & $3865$        & $635$         & $3750$        & $635$         \\
Var: Department:Discipline (Intercept) & $0.33$        & $0.32$        & $0.31$        & $0.19$        & $0.35$        & $0.47$        & $0.22$        & $0.13$        \\
\hline
\multicolumn{9}{l}{\scriptsize{$^{***}p<0.001$; $^{**}p<0.01$; $^{*}p<0.05$}}
\end{tabular}

\caption{As an alternate specification to the institution-discipline regressions, here we present the regression coefficients from a hierarchical Poisson regression on individual faculty.
    The model names correspond to the dependent variable (p=productivity, gp=group productivity, gs=group size) and the partition of the disciplines (c=disciplines with collaboration norms, nc=disciplines with no collaboration norms).
    We also include versions of the p/c model without the prestige and labor variables.
All continuous variables are standardized to have zero mean and unit variance.}
\label{tab:hierc_pois_reg}
\end{center}
\end{table*}

 %
\end{document}